\documentclass[preprint]{aastex}

\shorttitle{Vlasov Simulation}
\shortauthors{Yoshikawa et al.}

\newcommand{\itbold}[1]{\textbf{\textit{#1}}}
\newcommand{\pdif}[2]{\frac{\partial #1}{\partial #2}}

\begin{document}

\title{Direct integration of the collisionless Boltzmann
equation \\in six-dimensional phase space: Self-gravitating systems}

\author{Kohji Yoshikawa}
\affil{Center for Computational Sciences, University of Tsukuba, 1-1-1, 
Tennodai, Tsukuba, Ibaraki 305--8577, Japan}
\email{kohji@ccs.tsukuba.ac.jp}

\author{Naoki Yoshida}
\affil{Department of Physics, The University of Tokyo, 
Tokyo 113-0033, Japan}
\affil{Kavli Institute for the Physics and Mathematics of the Universe, 
The University of Tokyo, Kashiwa, Chiba 277-8583, Japan}

\and

\author{Masayuki Umemura}
\affil{Center for Computational Sciences, University of Tsukuba, 1-1-1, 
Tennodai, Tsukuba, Ibaraki 305--8577, Japan}

\begin{abstract}
  We present a scheme for numerical simulations of collisionless
  self-gravitating systems which directly integrates the
  Vlasov--Poisson equations in six-dimensional phase space. By the
  results from a suite of large-scale numerical simulations, we
  demonstrate that the present scheme can simulate collisionless
  self-gravitating systems properly. The integration scheme is based
  on the positive flux conservation method recently developed in
  plasma physics. We test the accuracy of our code by performing
  several test calculations including the stability of King spheres,
  the gravitational instability and the Landau damping.  We show that
  the mass and the energy are accurately conserved for all the test
  cases we study.  The results are in good agreement with linear
  theory predictions and/or analytic solutions.  The distribution
  function keeps the property of positivity and remains
  non-oscillatory.  The largest simulations are run on $64^{6}$ grids.
  The computation speed scales well with the number of processors, and
  thus our code performs efficiently on massively parallel
  supercomputers.
\end{abstract}

\keywords{galaxies: kinematics and dynamics --- methods: numerical}

\section{Introduction}
Gravitational interaction is one of the most important physical
processes in the dynamics and the formation of astrophysical 
objects such as star
clusters, galaxies, and the large scale structure
of the universe. Stars and dark matter in these self-gravitating systems
are essentially collisionless, except for a few cases such as
globular clusters and stars around supermassive blackholes. 
The dynamics of the collisionless systems is described by the
collisionless Boltzmann equation or the Vlasov equation.

Conventionally, gravitational $N$-body simulations are used to follow
the evolution of collisionless systems.  In such simulations,
particles represent sampled points of the distribution function in the
phase space.  The particles -- point masses -- interact
gravitationally with other particles, through which their orbits are
determined.  They are actually super-particles of stars or dark matter
particles. The gravitational potential field reproduced in a $N$-body
simulation is therefore intrinsically grainy rather than what it
should be in the real physical system.  It is well known that two-body
encounters can alter the distribution function in the way which
violate the collisionless feature of the systems, and undesired
artificial two-body relaxation is often seen in $N$-body simulations.
There is another inherent problem in $N$-body simulations.
Gravitational softening needs to be introduced to avoid artificial
large-angle scattering of particles caused by close encounters.
Physical quantities such as mass density and velocity field are
subject to intrinsic random noise owing to the finite number of
particles, especially in low-density regions.

To overcome these shortcomings of the $N$-body simulations, several
alternative approaches have been explored. For example, the
self-consistent field (SCF) method \citep{Hernquist1992, Hozumi1997}
integrates orbits of particles under the gravitational field
calculated by expanding the density and the gravitational potential
into a set of basis functions.  In the SCF method, the particles do
not directly interact with one another but move on the smooth
gravitational potential calculated from the overall distribution of
the particles. Despite of these attractive features, the major
disadvantage of the SCF method is its inflexibility that the basis set
must be chosen so that the lowest order terms reproduce the global
structure of the systems under investigation \citep{MWeinberg1999}.
In other words, the SCF method can be applied only to the secular
evolution of the collisionless systems.

The ultimate approach for numerical simulations of the collisionless
self-gravitating systems would be direct integration of the
collisionless Boltzmann equation, or Vlasov equation, combined with
the Poisson equation. The advantage of the Vlasov--Poisson simulations
was already shown by \citet{Janin1971} and \citet{Cuperman1971}, who
studied one-dimensional violent relaxation problems using the
water-bag method \citep{Hohl1967, Roberts1967}.
\citet{Fujiwara1981,Fujiwara1983}, for the first time, successfully
solved the Vlasov--Poisson equations for one-dimensional and
spherically symmetric systems using the finite volume method.  Other
grid-based approaches include the seminal splitting method of
\citet{Cheng1976}, more generally the semi-Lagrangean methods
\citep{Sonnendrucker1998}, a finite element method \citep{Zaki1988}, a
finite volume method \citep{Filbet2001}, the spectral method
\citep{Klimas1987, Klimas1994}, and a more recent multi-moment method
\citep{Minoshima2011}.  A comparison study of some of these methods is
presented in \citet{Filbet2003}.

So far, such direct integration of the Vlasov equation has been applied
only to problems in one or two spatial dimensions.  Solving the Vlasov
equation in six-dimensional phase space requires an extremely large
memory and computational time.  However, the rapid development of
massively parallel supercomputers has made it possible to simulate
collisionless self-gravitating systems in the full six-dimensional phase
space by numerically integrating the Vlasov--Poisson equations with a
scientifically meaningful resolution.

In this paper, we present the results from a suite of large simulations
of collisionless self-gravitating systems.  To this end, we develop a
fully parallelized Vlasov--Poisson solver.  We perform an array of test
calculations to examine the accuracy of our simulation code.  We compare
the obtained results with analytic solutions as well as linear theory
predictions.  We discuss the advantage and disadvantage of the
Vlasov--Poisson approach over the conventional $N$-body method.

The rest of the paper is organized as follows. Section 2 is devoted to
describe the detailed implementation of our numerical code to directly
integrate the Vlasov--Poisson equations. In section 3, we present the
results of several test runs and their comparison with those obtained
with the $N$-body method. The CPU timing and the parallelization
efficiency are presented in section 4. Finally, in section 5, we
summarize our results.

\section{Numerical Scheme}

For a collisionless self-gravitating system, the distribution function
of matter $f(\itbold{x},\itbold{v},t)$ obeys the Vlasov--Poisson equations
\begin{equation}
 \label{eq:vlasov}
 \pdif{f}{t}+\itbold{v}\cdot\pdif{f}{\itbold{x}} - \pdif{\phi}{\itbold{x}} \cdot \pdif{f}{\itbold{v}} = 0,
\end{equation}
where $\itbold{x}$ and $\itbold{v}$ are the spatial and velocity
coordinates, and $\phi$ is the gravitational potential satisfying the
Poisson equation
\begin{equation}
 \label{eq:poisson}
 \nabla^2\phi = 4\pi G\rho = 4\pi G\int f {\rm d}^3\itbold{v}.
\end{equation}
We normalize the distribution function so that its integration over
entire velocity space yields the mass density.

In order to numerically compute equations (\ref{eq:vlasov}) and
(\ref{eq:poisson}) simultaneously using the finite volume method, we
configure $N_{x}\times N_{y}\times N_{z}$ uniformly spaced Cartesian
grids (the spatial grids) in a simulation volume defined in $-L_{x}/2 <
x < L_{x}/2$, $-L_{y}/2 < y < L_{y}/2$, and $-L_{z}/2 < z < L_{z}/2$. We
also configure $N^{\rm v}_{x}\times N^{\rm v}_{y}\times N^{\rm v}_{z}$
uniform Cartesian grids (the velocity grids) in the velocity space with
$V_{x}^{-} < v_{x} < V_{x}^{+}$, $V_{y}^{-} < v_{y} < V_{y}^{+}$, and
$V_{z}^{-} < v_{z} < V_{z}^{+}$ at each spatial grid. Thus, the grid
spacings are given by
\begin{equation}
 \Delta x = \frac{L_x}{N_x},\,\,\,\,\Delta y = \frac{L_y}{N_y},\,\,\,\,\Delta z = \frac{L_z}{N_z}
\end{equation}
and 
\begin{equation}
 \Delta v_x = \frac{V^{+}_x-V^{-}_x}{N^{\rm v}_x},\,\,\,\,\Delta v_y = \frac{V^{+}_y-V^{-}_y}{N^{\rm v}_y},\,\,\,\,\Delta v_z = \frac{V^{+}_z-V^{-}_z}{N^{\rm v}_z}
\end{equation}
for the spatial and velocity grids, respectively.

\subsection{Vlasov Solver}

We adopt the time splitting scheme proposed by \citet{Cheng1976}.  The
Vlasov equation is split into one-dimensional advection equations for
each dimension of the phase space. Practically, we solve the following
six one-dimensional advection equations sequentially; three for the
advection in the position space
\begin{equation}
 \label{eq:adv_xpos}
 \pdif{f}{t} + v_x\pdif{f}{x} = 0
\end{equation}
\begin{equation}
 \label{eq:adv_ypos}
 \pdif{f}{t} + v_y\pdif{f}{y} = 0
\end{equation}
\begin{equation}
 \label{eq:adv_zpos}
 \pdif{f}{t} + v_z\pdif{f}{z} = 0 
\end{equation}
and the remaining three equations in the velocity space
\begin{equation}
 \label{eq:adv_xvel}
 \pdif{f}{t} - \pdif{\phi}{x}\pdif{f}{v_x} = 0
\end{equation}
\begin{equation}
 \label{eq:adv_yvel}
 \pdif{f}{t} - \pdif{\phi}{y}\pdif{f}{v_y} = 0
\end{equation}
\begin{equation}
 \label{eq:adv_zvel}
 \pdif{f}{t} - \pdif{\phi}{z}\pdif{f}{v_z} = 0.
\end{equation}

A number of schemes are available to solve the advection equations on
regular grids, such as the semi-Lagrange scheme
\citep{Cheng1976,Sonnendrucker1998} and the spectral method
\citep{Klimas1987, Klimas1994}.  An important property of the Vlasov
equation is the conservation of the phase space density of matter,
which leads to the conservation of mass in the system. Therefore, it
is quite natural to adopt a manifestly conservative scheme. Also the
positivity of the phase space density has to be ensured. In this
paper, we adopt the Positive Flux Conservation (PFC) scheme proposed
by \citet{Filbet2001} for the time evolution of the advection
equation.  The PFC scheme, by construction, ensures the conservation
of the mass, the preservation of the positivity, and the maximum
principle.

Here, we describe the PFC scheme briefly. Let us consider 
discretizing the following one-dimensional advection equation
\begin{equation}
 \label{eq:1d_adv}
 \pdif{f(x,t)}{t} + u \pdif{f(x,t)}{x} = 0. 
\end{equation}
Let $f^n_i$ be the averaged value of the distribution function at a
spatial region with the central value of $x_i$ 
and the interval of $\Delta x$ such that
\begin{equation}
 f^n_i \Delta x = \int_{x_i - \Delta x/2}^{x_i+\Delta x/2}f(x,t^n)\, {\rm d}x.
\end{equation}
Suppose the values of the distribution function $f_i^n$ at a
time of $t^n=n\Delta t$ are known for a finite set of grid points.  The
conservation of the phase space density leads to
\begin{equation}
 \label{eq:conserve}
 \int_{x_i-\Delta x/2}^{x_i+\Delta x/2} f(x,t^{n+1}) \,{\rm d}x = \int_{X(t^n,t^{n+1}, x_i-\Delta x/2)}^{X(t^n,t^{n+1}, x_i+\Delta x/2)}f(x,t^n) \,{\rm d}x,
\end{equation}
where $t^{n+1} = t^n + \Delta t$ and $X(t_1, t_2,x)$ is the value of the
$x$-coordinate of the characteristic curve at a time of $t=t_1$
originating from the phase space coordinate $(t_2,x)$. By denoting 
\begin{equation}
 \Phi^+ = \frac{1}{\Delta x}\int_{X(t^n,t^{n+1}, x_i+\Delta x/2)}^{x_i+\Delta x/2} f(x,t^n)\,{\rm d}x
  \label{eq:phi_plus}
\end{equation}
and
\begin{equation}
 \Phi^- = \frac{1}{\Delta x}\int_{X(t^n,t^{n+1}, x_i-\Delta x/2)}^{x_i-\Delta x/2} f(x,t^n)\,{\rm d}x,
  \label{eq:phi_minus}
\end{equation}
Equation (\ref{eq:conserve}) can be rewritten as 
\begin{equation}
 f^{n+1}_i = f^n_i + \Phi^{-} - \Phi^{+}.
\end{equation}
We compute $\Phi^+$ and $\Phi^-$ by interpolating the values of the
distribution function at the grid points. Specifically, we adopt the
third order approximation of $f(x,t^n)$ with a slope corrector to
suppress artificial numerical oscillations \citep{Filbet2001}.  As for
the boundary condition in solving the one-dimensional advection
equations, the outflow boundary condition is implemented in the
velocity space. Thus, when the matter is accelerated beyond the
predefined velocity limit $V^{\pm}_{x,y,z}$, it is regarded as
vanished. In the position space, both of the periodic and outflow
boundary conditions are available depending on problems.

Using the PFC scheme for the numerical integration of one-dimensional
advection equations, we advance of the distribution
function from $f(\itbold{x}, \itbold{v}, t^n)$ to $f(\itbold{x},
\itbold{v}, t^{n+1})$ by sequentially updating each one-dimensional
advection equation as
\begin{eqnarray}
 \displaystyle f(\itbold{x},\itbold{v},t^{n+1}) =
& T_{v_z}(\Delta t/2)T_{v_y}(\Delta t/2)T_{v_x}(\Delta t/2)& \nonumber \\ &T_x(\Delta t)T_y(\Delta t)T_z(\Delta t)&\nonumber \\
 &T_{v_z}(\Delta t/2)T_{v_y}(\Delta t/2)T_{v_x}(\Delta
  t/2)f(\itbold{x},\itbold{v},t^n), 
  \label{eq:integration}
\end{eqnarray}
where $T_l(\Delta t)$ denotes the numerical advection operator along
$l$-direction for a timestep of $\Delta t$. Here, we solve the Poisson
equation after operating the advection equations in the position space.
This time integration scheme is equivalent to the second
order leapfrog scheme.

\subsection{Poisson Solver}

The gravitational potential $\phi$ is computed under the periodic boundary
conditions or the isolated boundary conditions. For a given distribution
function $f(\itbold{x}, \itbold{v}, t)$, the mass density $\rho$ at a
spatial grid point $\itbold{x}$ is obtained simply by integrating the
distribution function over the velocity space,
\begin{equation}
 \rho(\itbold{x}) = \int f(\itbold{x}, \itbold{v}, t)\,{\rm d}^3 \itbold{v}.
\end{equation}
We adopt the convolution method with the
Fourier transform \citep{Hockney1981} to numerically solve the Poisson
equation.

For the periodic boundary conditions, we first compute the discrete
Fourier transform (DFT) of the density $\hat{\rho}(\itbold{k})$ using
the fast Fourier transform (FFT), where $\itbold{k}=(k_x, k_y, k_z)$
is a wave-number vector. Then, the Fourier-transformed gravitational
potential is given by
\begin{equation}
 \hat{\phi}(\itbold{k}) = \hat{G}(\itbold{k})\hat{\rho}(\itbold{k}),
\end{equation}
where $\hat{G}(\itbold{k})$ is the DFT of the green function of the
discretized Poisson equation. For $\Delta x = \Delta y =
\Delta z = \Delta$, it is given by
\begin{equation}
 G(\itbold{k}) = -\frac{\pi G \Delta^2}{\sin^2(k_x \Delta/2)+\sin^2(k_y\Delta/2)+\sin^2(k_z\Delta/2)}.
\end{equation}
Finally, the inverse FFT of $\hat{\phi}(\itbold{k})$ yields the
gravitational potential $\phi(\itbold{x})$ in the real space. 

As for the isolated boundary condition, we adopt the doubling up method
\citep{Hockney1981}, in which the number of the spatial grid points is
doubled for all coordinate axes, and the mass densities in the extended
grid points are set to zero. The Green function is constructed as
follows. First, it is defined at $N_x\times N_y\times N_z$ grid points
in real space as
\begin{equation}
   G(x,y,z) = \frac{G}{(x^2+y^2+z^2)^{1/2}}
\end{equation}
for $0\le x\le L_x$, $0\le y\le L_y$, $0\le z\le L_z$. By duplicating
and mirroring it in the extended grid points, we obtain the Green
function periodic in the $2N_{\rm x}\times 2N_{\rm y} \times 2N_{\rm z}$
grid points. After computing the Fourier transform of the Green function
$\hat{G}(\itbold{k})$, the gravitational potential in the real space
$\phi(\itbold{x})$ is obtained in the same manner as in the periodic
boundary condition.

In order to calculate the gravitational force at each spatial grid point,
we adopt the 2-point finite-difference scheme, 
in which the
gradient of the gravitational potential is calculated as 
\begin{equation}
\left(\frac{\phi_{i+1,j,k}-\phi_{i-1,j,k}}{2\Delta x}, \frac{\phi_{i,j+1,k}-\phi_{i,j-1,k}}{2\Delta y},\frac{\phi_{i,j,k+1}-\phi_{i,j,k-1}}{2\Delta z}\right), 
\end{equation}
where $\phi_{i,j,k}$ is the gravitational potential at a spatial grid
point with indices of $(i,j,k)$.

\subsection{Parallelization}

The computational cost for the time integration of the Vlasov equation
and the required amount of memory to store the distribution function in
the phase space roughly scale proportional to $N_x N_y N_z \times N^{\rm
v}_x N^{\rm v}_y N^{\rm v}_z$.  Hence efficient parallelization is
indispensable for the numerical integration of the Vlasov--Poisson
equations.

To parallelize our Vlasov--Poisson solver, we decompose the computational
domain in the phase space as follows.  The position space is divided
along each spatial axes into subdomains, while the velocity space at a
given spatial position is not decomposed.  In this way we can achieve an
equal balance in memory on distributed memory computers.  We use the
Message Passing Interface (MPI) for the inter-node parallelization; each
MPI process operates on a decomposed phase space. We also use the OpenMP
implementation to utilize the multi-thread parallelization on multi
CPU-cores in individual nodes. To solve the advection equations along
the spatial coordinate ($x$-, $y$- and $z$-coordinate) on each MPI
process, the values of the distribution function at the adjacent spatial
grid points are exchanged between the computational nodes. In solving
the Poisson equation, we do not parallelize the FFT because the required
computational cost of the FFT is nearly negligible compared with other
portions of the calculations and also because the size of FFT ($N_x N_y
N_z$) is not large enough for sufficient speed-up of the calculations.

\subsection{Timestep}

In solving the one-dimensional advection equation (\ref{eq:1d_adv})
using the PFC method described in section 2, the timestep width
$\Delta t$ is not restricted by the Courant--Friedrichs--Levy (CFL)
condition. However, when integrating the multi-dimensional
Vlasov--Poisson equations, we need to constrain the timestep under the
following considerations; (i) the accuracy of the characteristic lines
is better for the smaller $\Delta t$. (ii) to integrate the Vlasov
equation on a distributed memory system with phase space domain
decomposition, the exchange of the distribution function at the
boundaries of subdomains is unavoidable.  If we set too large a
timestep width, the trajectories of the characteristic lines get more
distant from the boundaries and then the number of grid points whose
data should be sent to the adjacent subdomains becomes also larger,
resulting in the increase of data exchange among the MPI processes.

We constrain the timestep for integrating the Vlasov equation as
\begin{equation}
 \Delta t =  C \min(\Delta t_{\rm p}, \Delta t_{\rm v}),
\end{equation}
where $\Delta t_{\rm p}$ and $\Delta t_{\rm v}$ is the timestep
constraints for the advection equations in position space
(\ref{eq:adv_xpos})--(\ref{eq:adv_zpos}) given by
\begin{equation}
 \Delta t_{\rm p} = \min\left(\frac{\Delta x}{V_x^{\rm max}}, \frac{\Delta y}{V_y^{\rm max}},\frac{\Delta z}{V_z^{\rm max}}\right)
\end{equation}
and for the ones in velocity space
(\ref{eq:adv_xvel})--(\ref{eq:adv_zvel}) given by
\begin{equation}
 \Delta t_{\rm v} = \min_i\left(\frac{\Delta v_x}{|a_{x,i}|},\frac{\Delta v_y}{|a_{y,i}|},\frac{\Delta v_z}{|a_{z,i}|}\right),
\end{equation}
where $a_{x,i}$, $a_{y,i}$ and $a_{z,i}$ are the $x$, $y$ and
$z$-components of the gravitational acceleration, $\nabla\phi$, at the
$i$-th grid point, and the minimization is taken over all the spatial
grids.

\section{Test Calculations}

In this section, we present a series of Vlasov--Poisson simulations of
self-gravitating systems using our newly developed parallel code.

\subsection{Test 1: 1-Dimensional Advection}

As a test of the PFC scheme to solve a 1-dimensional advection
equation, we perform simulations of 1-dimensional freely streaming
matter. This is the most trivial test, but it is indeed important to
check the positivity and non-oscillatory behaviour of the distribution
function. We solve the following Vlasov equation without the
gravitational acceleration term,
\begin{equation}
 \pdif{f}{t} + v\pdif{f}{x}=0.
\end{equation}
Here, we consider a 2-dimensional phase space defined as
\begin{equation}
 \left\{
 \begin{array}{c}
  -L/2 \le  x  \le L/2 \\
  -V_{\rm m} \le v\le V_{\rm m}
 \end{array},
 \right.
\end{equation}
where we impose the periodic boundary condition for the
$x$-coordinate. The initial condition is given by
\begin{equation}
 \left\{
  \begin{array}{cc}
   f(x,v,t=0) = 1 & -L/4\le x \le L/4 \mbox{ and } -V_{\rm m}/2 \le v \le V_{\rm m}/2 \\
   f(x,v,t=0) = 0 &  \mbox{otherwise}, \\
  \end{array}
 \right.
\end{equation}
It is expected that for each velocity $v$ the distribution function is
translated with time at a speed of $v$ and its shape with respect to
$x$ is preserved. The numbers of grids along $x$- and $v$-coordinates
are both set to be 128.

\begin{figure}[htbp]
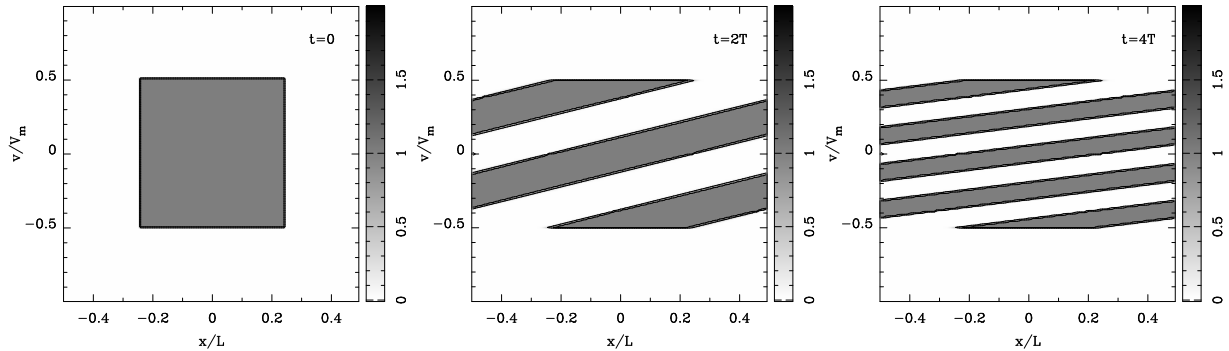

 \leavevmode
 \begin{center}
  \includegraphics[keepaspectratio,width=5.3cm]{pfc_adv_0.0.eps}
  \includegraphics[keepaspectratio,width=5.3cm]{pfc_adv_2.0.eps}
  \includegraphics[keepaspectratio,width=5.3cm]{pfc_adv_4.0.eps}
  \figcaption{Test 1: The phase space density of one-dimensional
  free-streaming matter at $t=0$ (left), $2T$ (middle) and $4T$
  (right). Black lines show the contours for $f(x,v,t)=0.5$ and
  1.0. \label{fig:1d_adv}}
 \end{center}
\end{figure}

Figure~\ref{fig:1d_adv} shows the phase space density at $t=0$, $2T$
and $4T$, where the system's unit $T$ is defined as $T\equiv
L/V_{\rm}$.  The black lines show the contour for $f(x,v,t)=0.5$ and
1.0.  We clearly see that the sharp edge of the phase space density is
well reproduced. We confirmed that there is no numerical oscillations
around the sharp edge and also that the distribution function is
always positive in the phase space.  

\begin{figure}[htbp]
 \leavevmode
  \begin{center}
   \includegraphics[keepaspectratio,width=10cm]{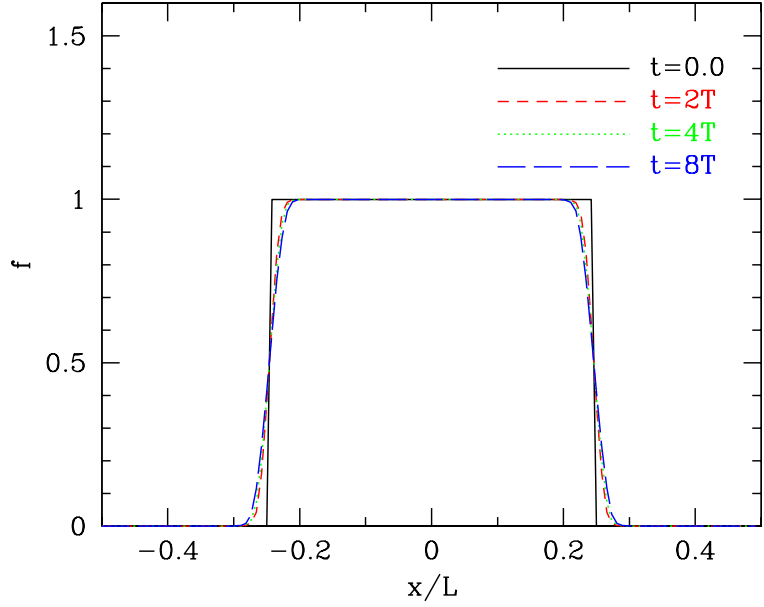}
   \figcaption{Test 1: Profiles of the distribution function along
   $v=0.5V_{\rm m}$ at $t=0.0$, $2T$, $4T$, and
   $8T$.\label{fig:1d_adv_profile}}
  \end{center}
\end{figure}

Profiles of the distribution function with respect to $x$ along
$v=V_{\rm m}/2$ at $t=0$, $2T$, $4T$ and $8T$ are shown in
\ref{fig:1d_adv_profile}. Since we impose the periodic boundary
condition in the $x$-direction, it is expected that the profile of the
distribution function remains the same at $t=0$, $2T$, $4T$ and $8T$
along $v=V_{\rm m}/2$. Although the sharp edges around $x=\pm L/4$ are
slightly smeared due to the phase error caused by numerical diffusion,
the profiles at $t=4T$ and $8T$ are almost the same.  Numerical
diffusion smears the distribution function only initially, but does
not cause secular errors.

\begin{figure}[htbp]
 \leavevmode
 \begin{center}
  \includegraphics[keepaspectratio,width=10cm]{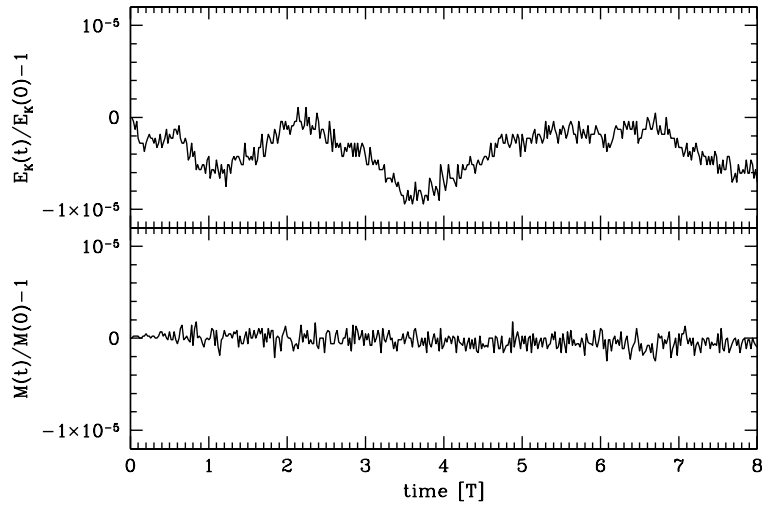}
  \figcaption{Test 1: The relative errors of the kinetic energy (upper
  panel) and the mass (lower panel). \label{fig:1d_adv_diag}}
 \end{center}
\end{figure}

Figure~\ref{fig:1d_adv_diag}
shows the relative errors of the kinetic energy $K(t)$ given by
\begin{equation}
 K(t) = \frac{1}{2}\int\int f(x,v,t)v^2 \,{\rm d}v{\rm d}x, 
  \label{eq:kinetic_energy}
\end{equation}
and the total mass $M(t)$
\begin{equation}
 M(t) = \int\int f(x,v,t) \,{\rm d}v{\rm d}x
  \label{eq:total_mass}
\end{equation}
during the calculation, manifesting that both of the kinetic energy and
the mass are conserved within the accuracy of $10^{-5}$. Note that the
PFC scheme for the 1-dimensional advection equation ensures the
conservation of the mass, the zeroth-order velocity moment of the
distribution function, but not the first- and second-order moment, and
that the conservation of the latters mainly depends on the numerical
resolution of the velocity space.

\subsection{Test 2: 1-Dimensional Homogeneous Self-Gravitating System}

In this test, we simulate a one-dimensional infinite self-gravitating
system following the Vlasov equation
\begin{equation}
 \pdif{f}{t} + v\pdif{f}{x} - \pdif{\phi}{x}\pdif{f}{v} = 0,
\end{equation}
coupled with the Poisson equation
\begin{equation}
 \nabla^2\phi = 4\pi G \rho = 4\pi G\int_{-\infty}^{\infty} f \,{\rm d}v,
\end{equation}
under the periodic boundary conditions in $x$-direction. We consider a
Maxwellian system with a periodic density fluctuation. The initial
distribution function is set to be
\begin{equation}
 f(x,v,t=0) = \frac{\bar{\rho}}{(2\pi\sigma^2)^{1/2}}\exp\left(-\frac{v^2}{2\sigma^2}\right)(1+A\cos kx),
\end{equation}
where $\bar{\rho}$ is the mean mass density, $\sigma$ is the velocity
dispersion and $A$ is the amplitude of the density fluctuation. In this
system, when the wave number of the density fluctuation $k$ is smaller
than the critical Jeans wave number $k_{\rm J}$ given by
\begin{equation}
 \label{eq:jeans_wavenumber}
 k_{\rm J} = \left(\frac{4\pi G\bar{\rho}}{\sigma^2}\right)^{1/2},
\end{equation}
the density fluctuation grows through the Jeans instability. On the
other hand, when $k>k_{\rm J}$, the density fluctuation damps through
the collisionless damping, or the Landau damping.

The computational domain of the two-dimensional phase space is set to be
\begin{equation}
 \left\{
 \begin{array}{c}
  -L/2 \le  x  \le L/2 \\
  -V \le v\le V 
 \end{array}
 \right. ,
\end{equation}
where $V$ is defined as $V = L/T$ and $T$ is the dynamical time defined by
\begin{equation}
 T = (G\bar{\rho})^{-1/2}.
\end{equation}
The number of grid points 
is 128 in both $x-$ and $v-$direction 
unless otherwise stated.

\begin{figure}[htbp]
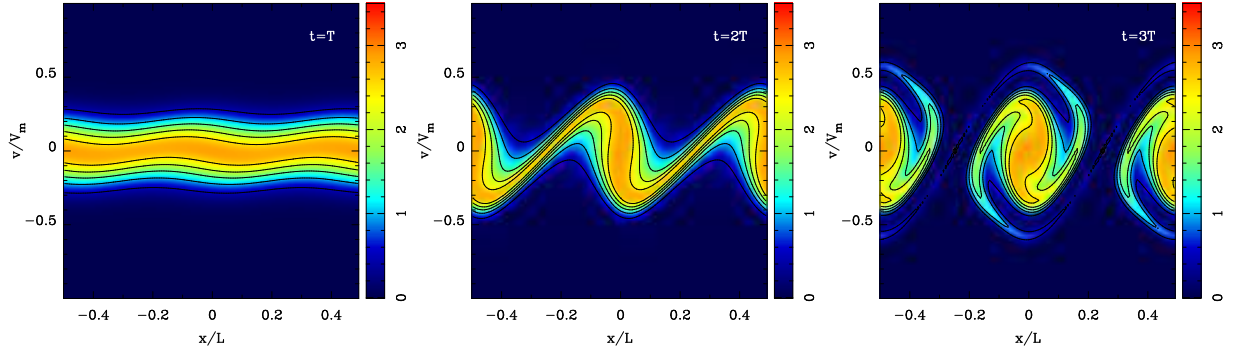

 \leavevmode
 \begin{center}
  \includegraphics[keepaspectratio,width=5.3cm]{vlasov_1d_koverkj_0.5_1.0.eps}
  \includegraphics[keepaspectratio,width=5.3cm]{vlasov_1d_koverkj_0.5_2.0.eps}
  \includegraphics[keepaspectratio,width=5.3cm]{vlasov_1d_koverkj_0.5_3.0.eps}
  \figcaption{Test 2: Phase space density in the run with $k/k_{\rm
  J}=0.5$ at $t=1.0\,T$ (left), $2.0\,T$ (middle) and $3.0\,T$
  (right). \label{fig:1d_grav}}
  \end{center}
\end{figure}

Since we impose the periodic boundary conditions, the wave number must
be set to $k=nk_0$, where $k_0=2\pi/L$ and $n$ is a positive integer,
and the velocity dispersion $\sigma$ is determined such that the ratio
$k/k_{\rm J}$ is adjusted to have a specific value.  In what follows,
the wave number is fixed to $k=2k_0$ ($n=2$). We show the results for
$k/k_{\rm J}=0.1$, 0.5, 1.1 and 2.0. The amplitude of the initial
density perturbation $A$ is set to $A=0.1$ for $k/k_{\rm J}>1$ and
$A=0.01$ for $k/k_{\rm J}<1$. Figure~\ref{fig:1d_grav} shows the phase
space density for the case with $k/k_{\rm J}=0.5$ at $t=T$, $2\,T$ and
$3\,T$. In this case, as expected, the density fluctuation grows
monotonically, and collapsed objects are formed through the
gravitational instability. Contrastingly, the density fluctuation is
damped through the Landau damping in the run with $k/k_{\rm J}=1.1$,
as can be seen in Figure~\ref{fig:1d_grav_2}.

\begin{figure}[htbp]
 \leavevmode
 \begin{center}
  \includegraphics[width=12cm]{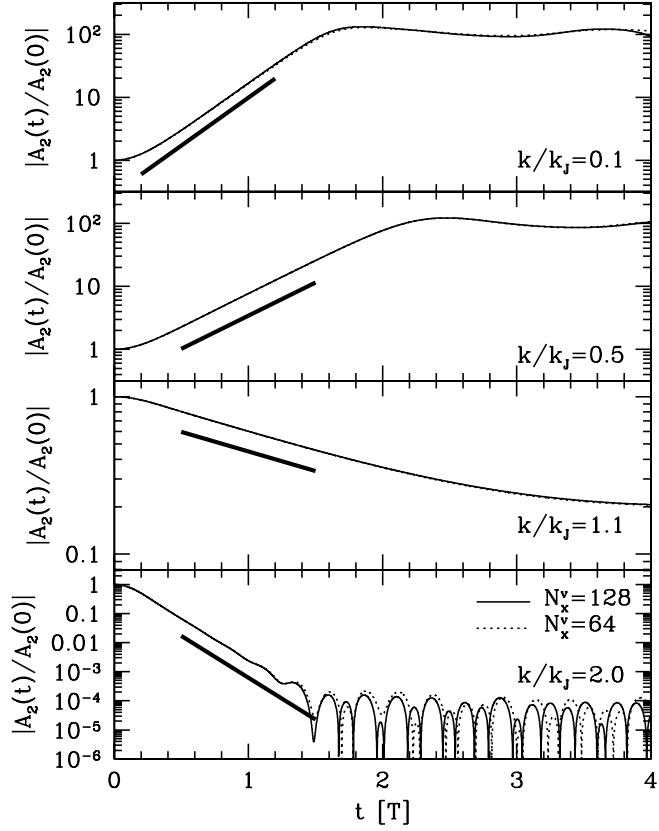} 
  \figcaption{Test 2: Time evolution of the density contrast at the
    density peak for the runs with $k/k_{\rm J}=0.1$, 0.5, 1.1 and 2.0
    (from top to bottom). The solid and dotted lines indicate the
    results with $N^{\rm v}_x=128$ and 64, respectively. The bold
    lines show the linear damping rate (see
    text).\label{fig:1d_grav_delta}}
 \end{center}
\end{figure}

Figure~\ref{fig:1d_grav_delta} shows the time evolution of the
amplitude of the density fluctuation $\delta \equiv
(\rho-\bar{\rho})/\bar{\rho}$ for $k/k_{\rm J}=0.1$ , 0.5, 1.1 and 2.0,
where the amplitude is expressed in terms of the Fourier amplitude $A_n$
which is given by
\begin{equation}
 \delta(x,t) = \sum_{n\ge0} A_n(t) \exp\left(i nk_0x\right).
\end{equation}
The time evolution of $|A_2(t)|$ is plotted in
Figure~\ref{fig:1d_grav_delta}. We also check the convergence of the
solution by doubling the resolution in the velocity space.  The results
from the runs with $N^{\rm v}_x=64$ and $N^{\rm v}_x=128$ are also
compared in Figure~\ref{fig:1d_grav_delta}.

The linear growth (or damping) rate $\gamma$ can be computed using the
dispersion relation
\begin{equation}
\label{eq:dispersion}
\frac{k^2}{k_{\rm J}^2} = 1+wZ(w),
\end{equation}
where $Z(w)$ is the plasma dispersion function
\begin{equation}
Z(w) = \frac{1}{\sqrt{\pi}}\int_{-\infty}^{\infty} {\rm d}s\,\frac{e^{-s^2}}{s-w}
\end{equation}
and $w$ is given by
\begin{equation}
w = \frac{\pm i \gamma}{\sqrt{8\pi G\rho}(k/k_{\rm J})}.
\end{equation}
A more detailed description on the growth rate and the dispersion
relation can be found in Binney \& Tremaine (2008).  For a given value
of $k/k_{\rm J}$, the growth and damping rates can be computed by
solving equation~(\ref{eq:dispersion}).  The bold line in each panel
in Figure~\ref{fig:1d_grav_delta} indicates the theoretical linear
growth or damping rate $\gamma$.  Our numerical results are in good
agreement with the linear theory prediction in the early phase.  Also
there is no significant difference between the results with $N^{\rm
  v}_x=64$ and $128$, indicating excellent convergence.  It is
interesting that the growth of the perturbation saturates at $T > 1$
in the run with $k/k_{\rm J} < 1$.  For $k/k_{\rm J}=2.0$, the
timescale of the damping is shorter than the dynamical timescale.  A
significant fraction of the mass are trapped in the trough of the
gravitational potential, with the distribution function being peaked
around $v=0$. Since such a distribution function with a small velocity
dispersion cannot damp the density fluctuation efficiently via the
Landau damping, density fluctuations begin oscillating after the early
linear damping phase.  Similarly, fluctuation damping saturates at
$t\gtrsim 3T$ for $k/k_{\rm J}=1.1$.  Figure~\ref{fig:1d_grav_2} shows
that the phase space density in the run with $k/k_{\rm J}=1.1$ departs
from the initial Gaussian distribution. The distribution function is
more concentrated around $v=0$ at later times. These features are also
pointed out by \citet{Fujiwara1981}.

\begin{figure}[htbp]
 \leavevmode
 \begin{center}
  \includegraphics[keepaspectratio,width=12cm]{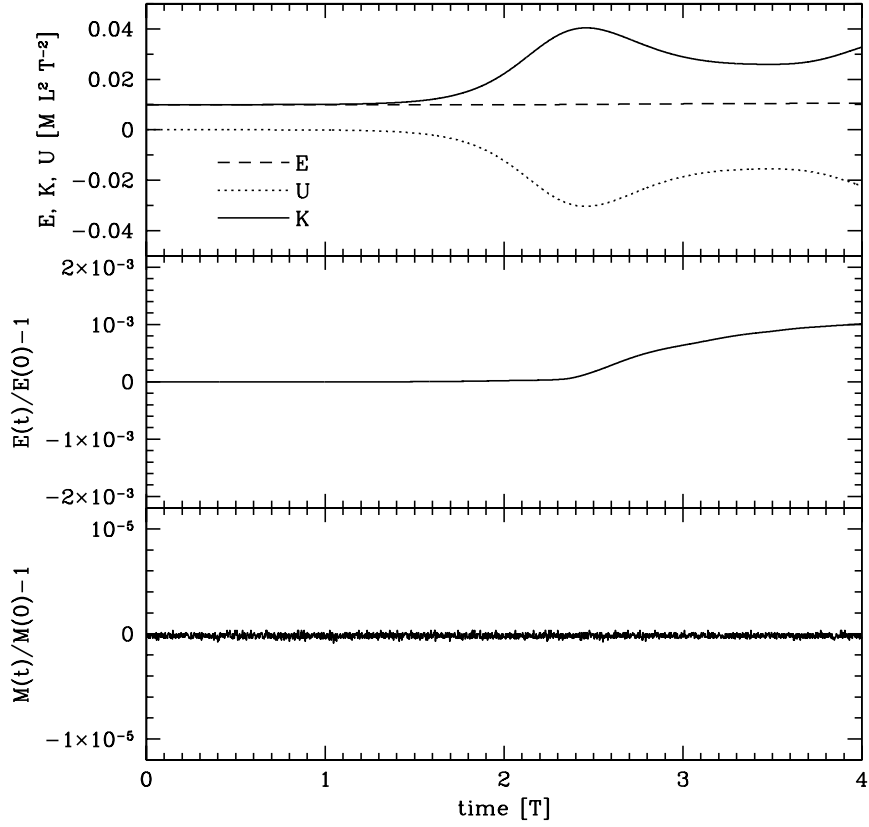}
  \figcaption{Test 2: The time evolutions of the kinetic, potential and
  total energy in the run with $k/k_{\rm J}=0.5$ are shown in the top
  panel. The relative errors of the total energy and the mass
  conservation are depicted in the middle and bottom panels,
  respectively. \label{fig:1d_grav_diag}}
 \end{center}
\end{figure}

The top panel of figure~\ref{fig:1d_grav_diag} shows the time evolution
of kinetic energy $K(t)$ given by equation~(\ref{eq:kinetic_energy}) and
the gravitational potential energy $U(t)$ computed as
\begin{equation}
 U(t) = \frac{1}{2}\int \rho(x) \phi(x) \,{\rm d}x,
\end{equation}
as well as the total energy $E = K(t) + U(t)$. The middle and
bottom panels indicate the relative errors in the
total energy $E$ and the total mass $M(t)$ given by
equation~(\ref{eq:total_mass}). The total energy and
the mass are conserved within the relative errors of $10^{-3}$ and
$10^{-6}$, respectively.

\begin{figure}[htbp]
 \leavevmode
 \begin{center}
  \includegraphics[keepaspectratio,width=5.3cm]{vlasov_1d_kovkj_1.1_1.0.eps}
  \includegraphics[keepaspectratio,width=5.3cm]{vlasov_1d_kovkj_1.1_2.0.eps}
  \includegraphics[keepaspectratio,width=5.3cm]{vlasov_1d_kovkj_1.1_4.0.eps}
  \figcaption{Test 2: Phase space density in the run with $k/k_{\rm
  J}=1.1$ at $t=1.0\,T$ (left), $2.0\,T$ (middle) and $4.0\,T$
  (right). \label{fig:1d_grav_2}}
  \end{center}
\end{figure}

\subsection{Test 3: Galilean Invariance}

It is well known that mesh-based hydrodynamical codes generally do not
assure the Galilean invariance because approximate Riemann solvers
employed in many of such codes are not manifestly Galilean invariant
\citep{Wadsley2008,Tasker2008}. This is in good contrast with
particle-based $N$-body simulations which are exactly Galilean invariant
as long as a symmetric time integration scheme is used. In the light of
this, it is interesting and important to examine the Galilean invariance
of our mesh-based scheme for self-gravitating systems. We test our code
by adding a constant translational velocity $v_{\rm t}$ to the initial
conditions of the Test 2 problem.  We compare the results with the
original one (with $v_{\rm t}=0$) presented in the previous section.

\begin{figure}[htbp]
 \leavevmode
 \begin{center}
  \includegraphics[width=13cm]{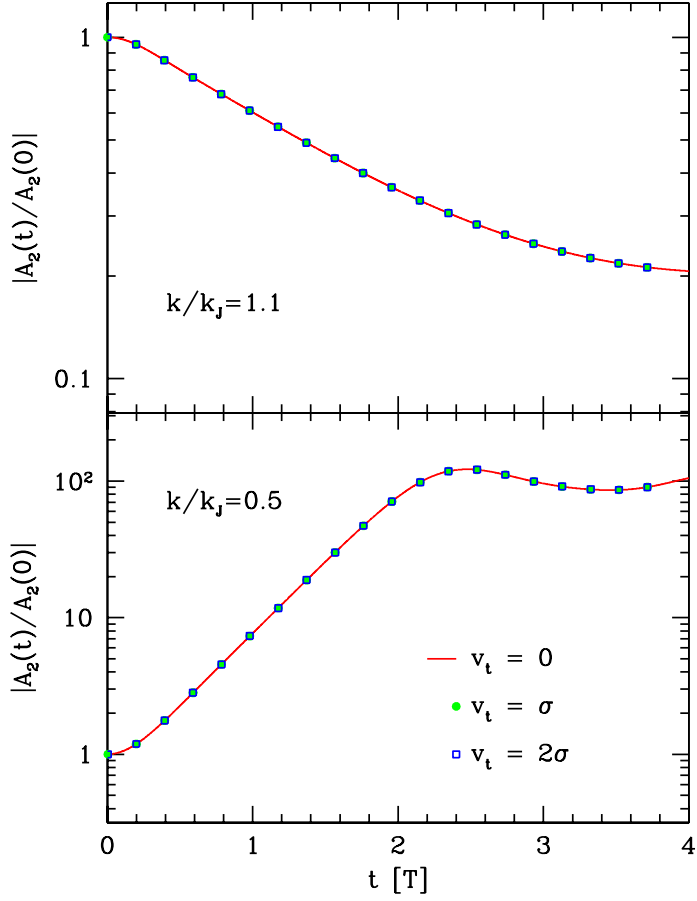}
  \figcaption{Test 3: Galilean invariance. We plot the time evolution
    of the amplitude of the density fluctuation of a mode of $n=2$
    with translational velocity of $v_{\rm t}=0$, $\sigma$ and
    $2\sigma$. Upper and lower panels show the results for $k/k_{\rm
      J}=1.1$ and $k/k_{\rm J}=0.5$, respectively.
 \label{fig:1d_grav_trans}}
 \end{center}
\end{figure}

Specifically, we set the initial distribution function as
\begin{equation}
 f(x, v, t=0) = \frac{\bar{\rho}}{(2\pi\sigma^2)^{1/2}}\exp\left(-\frac{(v-v_{\rm t})^2}{2\sigma^2}\right)(1+A\cos kx),
\end{equation}
where the velocity dispersion is set such that $k/k_{\rm J}=0.5$ and
$k/k_{\rm J}=1.1$ (see equation [\ref{eq:jeans_wavenumber}]).  We
assign $v_{\rm t}=\sigma$ and $2\sigma$. The numbers of the grid
points are set to $N_x=128$ and $N^{\rm v}_x=128$.
Figure~\ref{fig:1d_grav_trans} shows the comparison of the time
evolution of the density fluctuation of a $n=2$ mode with $v_{\rm
  t}=\sigma$ and $2\sigma$ to the original result with $v_{\rm t}=0$
presented in Test 2. Clearly, the results are independent of the
translational velocity and hence our code is Galilean invariant to
this accuracy.

\subsection{Test 4: 3-Dimensional Homogeneous Self-Gravitating System}

We study the gravitational instability and the Landau damping in a
3-dimensional self-gravitating system. We solve the Vlasov equation
coupled with the 3-dimensional Poisson equation in six-dimensional
phase space under the periodic boundary conditions for spatial
coordinates. The run is configured as follows. At each spatial grid,
the initial distribution function is given by
\begin{equation}
 f(\itbold{x},\itbold{v},t=0) = \frac{\bar{\rho}(1+\delta_i(\itbold{x}))}{(2\pi\sigma^2)^{3/2}}\exp\left(-\frac{|\itbold{v}|^2}{2\sigma^2}\right),
\end{equation}
where $\delta_i(\itbold{x})$ is the initial density fluctuation at a
spatial position $\itbold{x}$. The density fluctuations are generated by
assigning uniform random values between $-\delta_{\rm m}/2$ and
$\delta_{\rm m}/2$ so that the resulting density field has a white noise
power spectrum. We assign the velocity dispersion which is determined
from a predefined Jeans wavenumber $k_{\rm J}$.

The phase space volume with $-L/2 \le x,y,z \le L/2$ and $-V\le v_x,
v_y, v_z \le V$ is discretized with $N_x=N_y=N_z=64$ and $N^{\rm
v}_x=N^{\rm v}_y=N^{\rm v}_z=64$, where $V$ is again defined by $V =
L/T$ and $T = (G\bar{\rho})^{-1/2}$.  In what follows, $\delta_{\rm m}$
is set to 0.1, and we present the results for $k_{\rm
J}=2\pi/(L/4)=8\pi/L$ and $k_{\rm J}=2\pi/(L/8)=16\pi/L$. Note that the
velocity dispersions $\sigma$ in the runs with $k_{\rm J}=8\pi/L$ and
$k_{\rm J}=16\pi/L$ correspond to $9\Delta v_{x,y,z}$ and $4.5\Delta
v_{x,y,z}$, respectively. To characterize the three-dimensional density
fluctuations and their evolution, we compute the power spectrum 
$P(k)=\langle |\delta({\itbold k})|^2 \rangle$, where $\delta({\itbold
k})$ is the discrete Fourier transform of $\delta(\itbold{x})$ given by
\begin{equation}
 \delta(\itbold{x}) = \frac{1}{\bar{\rho}}\int f(\itbold{x},\itbold{v},t) \, {\rm d}^3\itbold{v} - 1.
\end{equation}

\begin{figure}[htbp]
 \leavevmode
 \begin{center}
  \includegraphics[keepaspectratio,width=13cm]{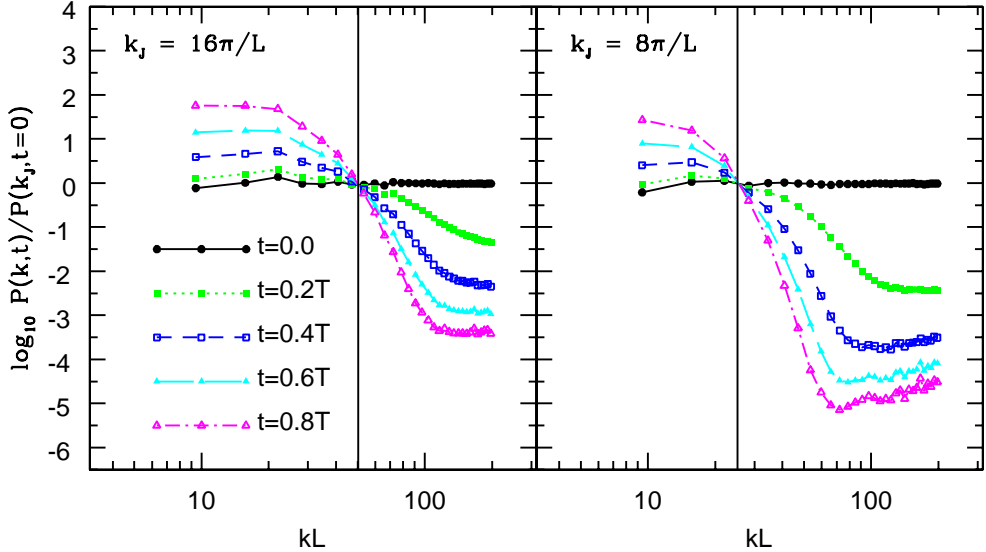}
  \figcaption{Test 4: Power spectra of the density fluctuation at
  $t/T=0.0$, 0.2, 0.4, 0.6 and 0.8 in the runs with $k_{\rm J}=8\pi/L$
  (right) and $k_{\rm J}=16\pi/L$ (left). The vertical line in each panel
  indicates the location of the Jeans
  wavenumber.\label{fig:3d_damping_pk}}
 \end{center}
\end{figure}

\begin{figure}[htbp]
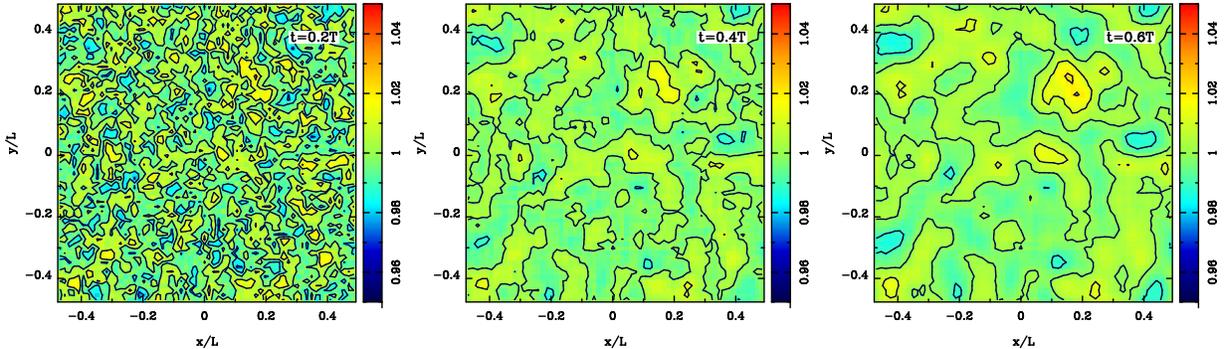

 \leavevmode
 \begin{center}
  \includegraphics[keepaspectratio,width=5.3cm]{damping_L8-03_dens.eps}
  \includegraphics[keepaspectratio,width=5.3cm]{damping_L8-05_dens.eps}
  \includegraphics[keepaspectratio,width=5.3cm]{damping_L8-07_dens.eps}

  \figcaption{Test 4: Maps of $\rho(\itbold{x})/\bar{\rho}$ on $z=0$
    planes at $t/T=0.2$, 0.4 and 0.6 in the run with $k_{\rm
      J}=16\pi/L$.  Contours are drawn
    for $0.96\le\rho(\itbold{x})/\bar{\rho}\le1.04$ with an interval of
    0.1.
  \label{fig:3d_damping_map}}
 \end{center}
\end{figure}

Figure~\ref{fig:3d_damping_pk} shows the power spectra of the density
field at $t=0$, $0.2T$, $0.4T$, $0.6T$, and $0.8T$ computed in the runs
with $k_{\rm J}=8\pi/L$ (left panel) and $k_{\rm J}=16\pi/L$ (right
panel), where the vertical line in each panel indicates the Jeans
wavenumber. As expected, the fluctuation modes with $k<k_{\rm J}$ grow
through the gravitational instability, whereas the modes with $k>k_{\rm
J}$ damp due to the Landau damping. These features can be directly
observed in the time evolution of the density fields in
$\itbold{x}$-space shown in figure~\ref{fig:3d_damping_map}, in which we
set $k_{\rm J}=16\pi/L$. We can see that density fluctuations with
smaller wavelength which is dominant at the early epoch ($t=0.2T$)
gradually vanish and only those with longer wavelength grow with time
through the gravitational instability. Note that the Jeans length in
this run ($k_{\rm J}=16\pi/L$) is $L/8$, and it can be seen, by visual
inspection, that there are no density fluctuations with wavelength much
smaller than the Jeans length $L/8$ at $t=0.6T$.

\begin{figure}[htbp]
 \leavevmode
 \begin{center}
  \includegraphics[keepaspectratio,width=13cm]{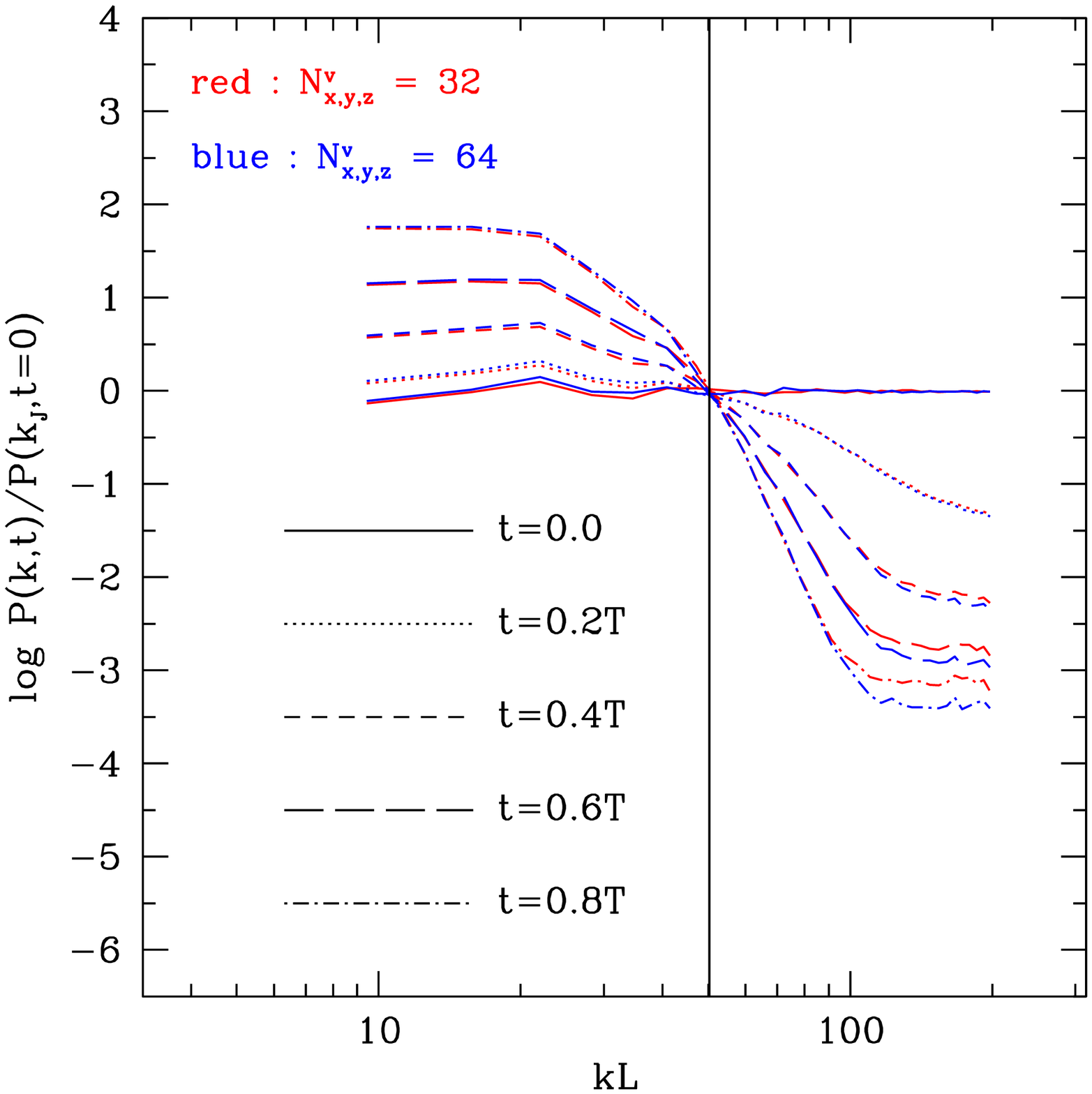}
  \figcaption{Test 4: The density power spectra at
  $t/T=0.0$, 0.2, 0.4, 0.6 and 0.8 in the runs with $N^{\rm
  v}_{x,y,z}=32$ (red) and $N^{\rm v}_{x,y,z}=64$ (blue) and with
  $k_{\rm J}=16\pi/L$. The vertical line indicates the location of the
  Jeans wavenumber.\label{fig:3d_damping_resolution}}
 \end{center}
\end{figure}

We have also performed a convergence test for the three-dimensional case
in order to examine the effect of the resolution in the velocity
space. We run the same simulation but with $N^{\rm v}_x=N^{\rm
v}_y=N^{\rm v}_z=32$ and $k_{\rm J}=16\pi/L$.  In
Figure~\ref{fig:3d_damping_resolution}, the resulting power spectra are
compared with those obtained in the run with $N^{\rm v}_x=N^{\rm
v}_y=N^{\rm v}_z=64$ for the same Jeans wave number. Overall an
excellent agreement is found except that the damping of modes of large
wave numbers $kL\gtrsim100$ is somewhat suppressed at late times in the
run with coarser velocity resolution.  Note that the vertical axis in
logarithmic scale; the amplitudes of the small-scale modes damp over
three orders of magnitude by $t=0.8T$.

\begin{figure}[htbp]
 \leavevmode
 \begin{center}
  \includegraphics[keepaspectratio,width=13cm]{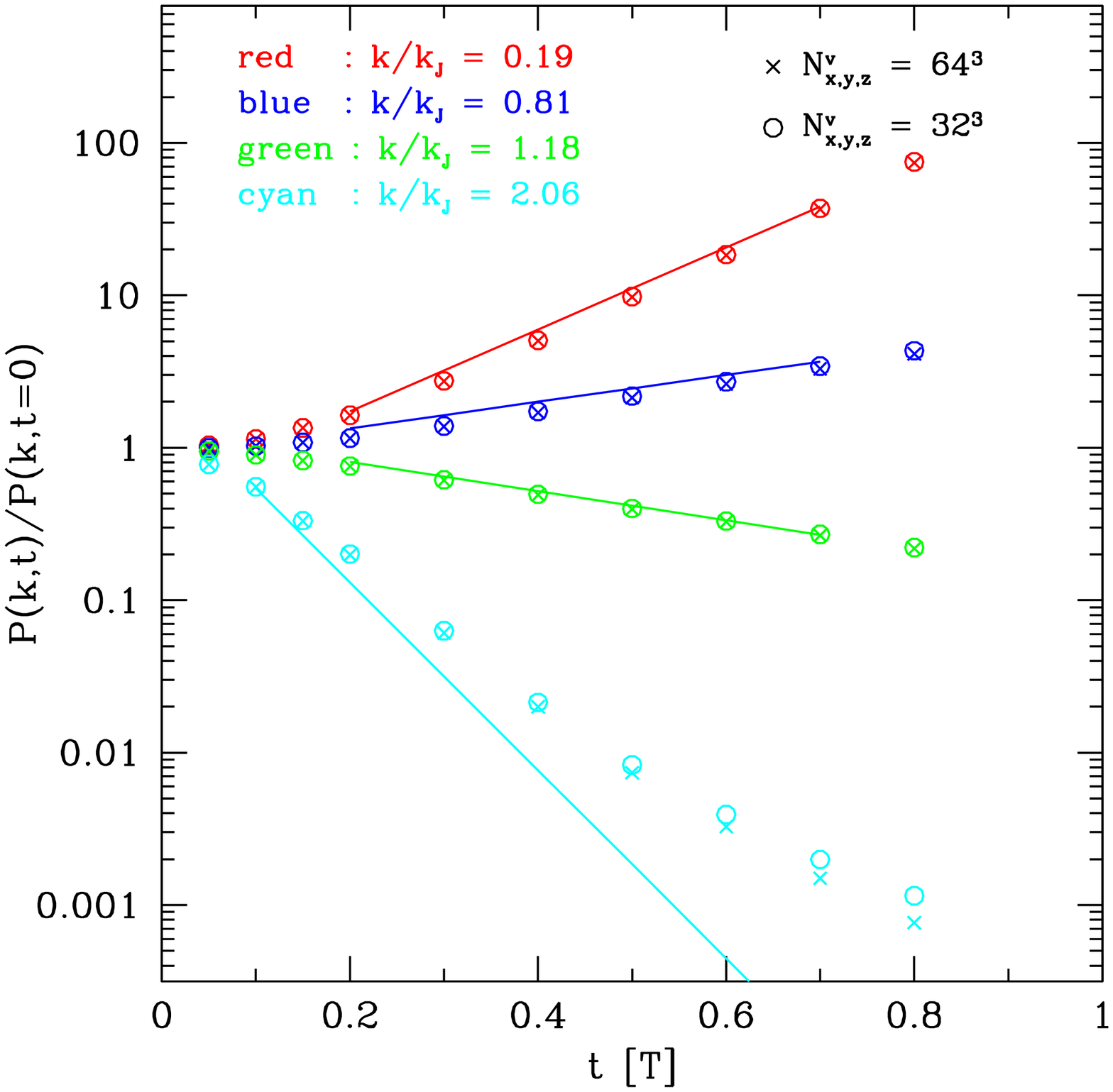}
  \figcaption{Test 4: Time evolution of $P(k)/P(k,t=0)$ for various
  values of $k/k_{\rm J}$. Results with $N^{\rm v}_x = N^{\rm v}_y =
  N^{\rm v}_z=64$ and $32$ are plotted.  Solid lines indicate the
  linear theory predictions,
  $P(k)/P(k,t=0)\propto \exp(2\gamma t)$, where $\gamma$ is the growth
  or damping rate of the density fluctuation. \label{fig:pk_ratio}}
 \end{center}
\end{figure}

Figure~\ref{fig:pk_ratio} shows the time evolution of the ratio between
the power spectra with respect to the initial power spectrum $t=0$,
$P(k,t)/P(k,t=0)$, for various $k/k_{\rm J}$. We plot both the results
with $N^{\rm v}_x = N^{\rm v}_y = N^{\rm v}_z=64$ and 32 for
comparison. In the linear regime, the ratios should be proportional to
$\exp(2\gamma t)$, where $\gamma$ is the growth or damping rate
calculated from the linear theory (see the previous section).  We show
the growth/damping of $\exp(2\gamma t)$ for the adopted value of
$k/k_{\rm J}$ as solid lines.  Figure~\ref{fig:3d_damping_resolution}
clearly shows that there is no significant difference between the
results with different velocity resolutions except for a very small
deviation at late times for a strongly damping mode of $k/k_{\rm
J}=2.08$.  It can be seen that the results with $k/k_{\rm J}=0.19$,
0.81, and 1.18 agree with the linear theory well. For $k/k_{\rm
J}=2.06$, however, the obtained damping rate is lower than the
theoretical prediction especially at $t\gtrsim 0.3T$. This feature is
also consistent with what we found in Test 2. It is likely owing to the
same mechanism of the suppression of Landau damping for a 'fluctuating'
mode with large $k/k_{\rm J}$ as discussed in Section 3.2.

\subsection{Test 5: King Sphere}

We perform a Vlasov--Poisson simulation of the King sphere. The
distribution function of the King sphere is a stable solution of the
Vlasov--Poisson equations and has a finite extension in the spatial
coordinate unlike other analytic stable solutions such as the Plummer
sphere and the Osipkov--Meritt model. The test is suitable for checking
the accuracy of the time integration of the Vlasov equation.

Let us denote the relative potential $\Psi(r)$ and relative energy
${\cal E}$ by
\begin{equation}
 \Psi(r) = -\Phi(r)
\end{equation}
and 
\begin{equation}
 {\cal E} = \Psi(r) - \frac{1}{2} (v_x^2 + v_y^2 + v_z^2),
\end{equation}
respectively, where $\Phi(r)$ is the gravitational potential with the
boundary condition $\Phi(r)\rightarrow 0$ as $r\rightarrow \infty$.
Then the distribution function of the King sphere is given by
\begin{equation}
 f({\cal E}) = \left\{
		\begin{array}{ll}
		 \rho_1(2\pi\sigma^2)^{-2/3}(e^{{\cal E}/\sigma^2}-1) & {\cal E} > 0 \\
		 0 & {\cal E} < 0
		\end{array},
	       \right.
\end{equation}
where $\rho_1$ and $\sigma$ are the constants which determine the
total mass $M$ and the overall shape of the King sphere. The shape of
the King sphere is characterized by the King parameter
$W=\Psi(0)/\sigma^2$, which we set $W=3$ in the followings. For $W=3$,
the tidal radius $r_t$, the outer boundary of the King sphere is $r_t
= 5.37r_0$, where $r_0 \equiv 3\sigma/\sqrt{4\pi G \rho_0}$ and
$\rho_0$ is the central mass density.

We consider the phase space volume with $-5.4 r_0 \le x, y, z\le 5.4
r_0$ and $ -1.5V \le v_x, v_y, v_z \le 1.5V$ where $V\equiv r_0/T$ and
$T = (GM/r_0^3)^{-1/2}$, and discretize it into grids with
$N_x=N_y=N_z=64$ and $N^{\rm v}_x=N^{\rm v}_y=N^{\rm v}_z=32$. In
setting up the initial condition, after the phase space density in
each phase space grid is calculated using the velocity and the
relative potential at the grid center, we re-normalize the total mass
of the King sphere so that the initial virial ratio $2K/|U|$ is unity,
where $K$ and $U$ is the total kinetic and gravitational potential
energy of the system.

\begin{figure}[htbp]
 \leavevmode
 \begin{center}
  \includegraphics[width=14cm]{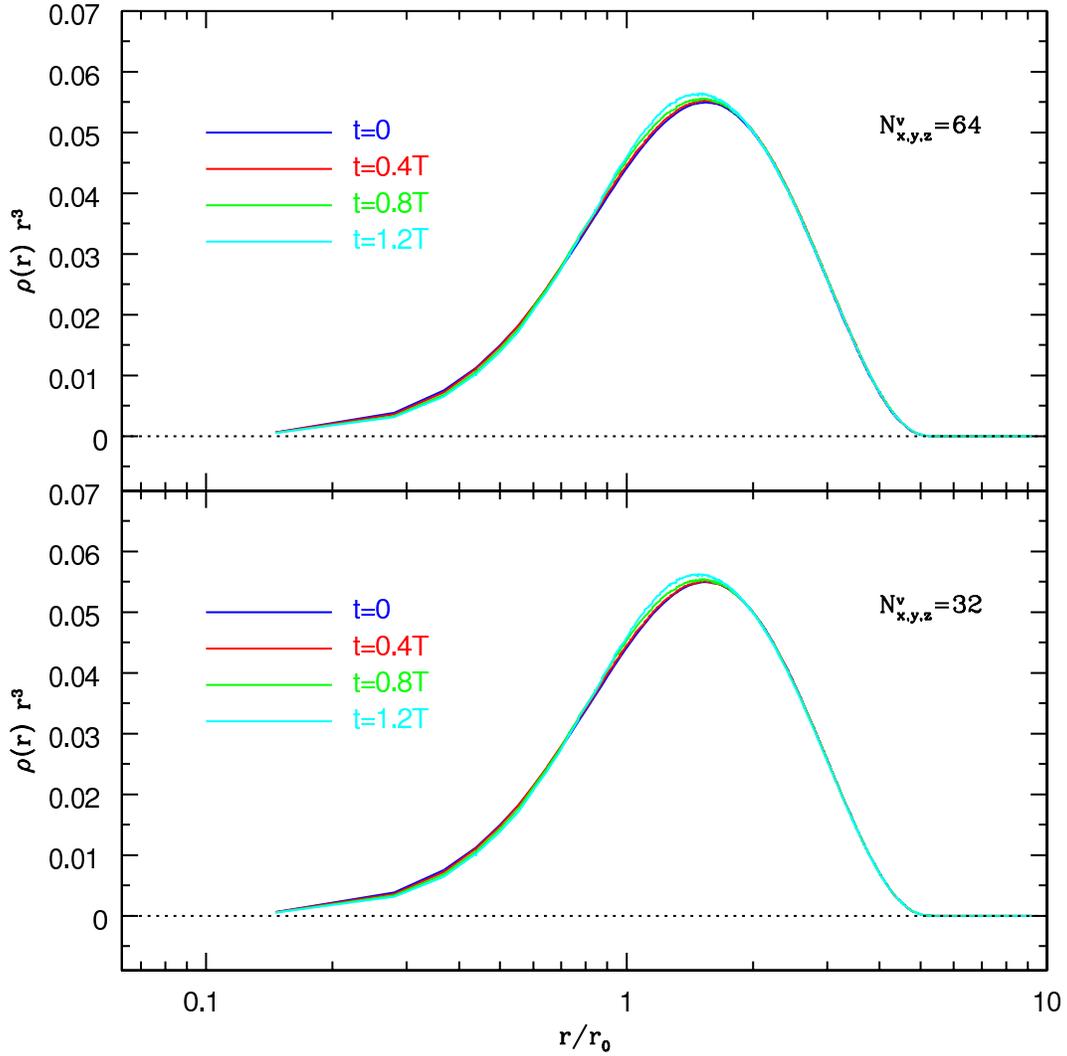} 
  \figcaption{Test 5: Mass distribution of the simulated King sphere at $t=0$,
  $0.4T$, $0.8T$ and $1.2T$. The numbers of the spatial grid points are
  set to $N_x = N_y = N_z = 64$, and those of the velocity grid are
  $N^{\rm v}_x=N^{\rm v}_y=N^{\rm v}_z=64$ (upper panel) and $N^{\rm
  v}_x=N^{\rm v}_y=N^{\rm v}_z=32$ (lower
  panel). \label{fig:king_sphere}}
 \end{center}
\end{figure}

Figure~\ref{fig:king_sphere} shows the mass distribution of the
simulated King sphere as a function of radius $r$. Note that, in this
figure, mass within a shell with $r_1 < r < r_2$ is proportional to
the area enclosed between the the profiles $\rho(r)r^3$ in $r_1 < r <
r_2$ and the horizontal dotted lines ($\rho(r)r^3=0$). It can be seen
that the mass distribution does not change significantly over one
dynamical timescale, irrespective of the numerical resolution of the
velocity space. In figure~\ref{fig:king_sphere}, we can see the slight
mass transfer from the inner part ($r\simeq 0.4r_0$) to the outer
($r\simeq 1.5r_0$) regions of the King sphere. Since the grid spacing
of the spatial grid is $\Delta x = \Delta y = \Delta z \simeq
0.0844r_0$ and the region with $r<r_0$ is resolved with only
$\simeq$10 grid points, the mass transfer can be ascribed to the
numerical diffusion of the PFC scheme seen in
Figure~\ref{fig:1d_adv_profile}.

\begin{figure}[htbp]
 \leavevmode
  \begin{center}
   \includegraphics[width=13cm]{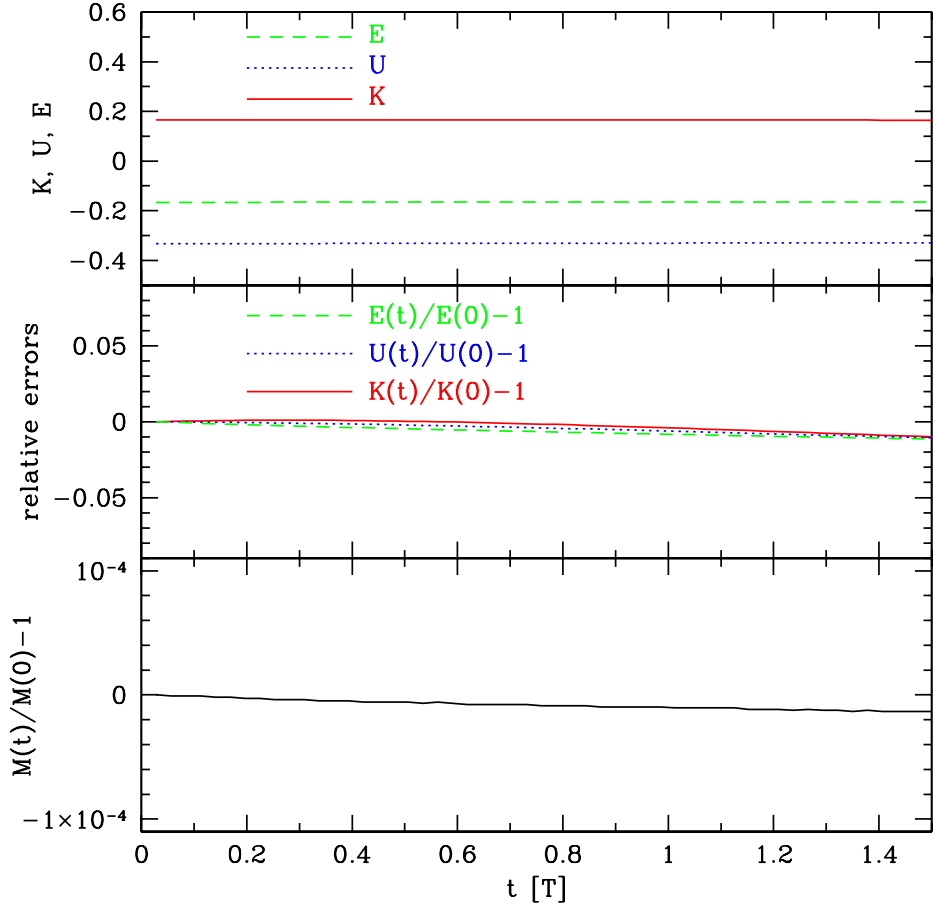} 

   \figcaption{Test 5: Time evolution of the kinetic, gravitational
   potential and total energy in the run with $N_x=N_y=N_z=N^{\rm
   v}_x=N^{\rm v}_y=N^{\rm v}_z=64$ and their relative difference are
   shown in the top and middle panels, respectively. The bottom panel
   depicts the relative error of the total
   mass. \label{fig:king_sphere_diag}}
  \end{center}
\end{figure}

Figure~\ref{fig:king_sphere_diag} shows the time evolution of the
kinetic, gravitational potential and total energy of the King sphere
simulated with $N_x=N_y=N_z=N^{\rm v}_x=N^{\rm v}_y=N^{\rm v}_z=64$.
Over one dynamical timescale, the total energy is kept constant with a
relative error of $\lesssim$ 1\%. The kinetic and gravitational
potential energies are also kept constant with good numerical accuracy
of 1\% as long as $t=T$. The total mass is also conserved with a
relative error of $\lesssim 10^{-4}$.

\subsection{Test 6: Merging of Two King Spheres}

As a final test, we perform a simulation of merging of two King
spheres using our Vlasov--Poisson solver.  We also run the same
simulation using a conventional $N$-body method and compare the
results.

The initial conditions are set up as follows. Two King spheres with
the same physical parameters as in Test-5 are initially located at
$(x,y,z)=(r_0, r_0, 0)$ and $(-r_0, -r_0, 0)$. The spheres are then
given bulk velocities of $-0.2V$ and $0.2V$ along $x$-axis,
respectively.  The phase space volume we consider has an dimension of
$-6.4r_0 \le x, y, z \le 6.4r_0$ and $-2.0V \le v_x, v_y, v_z \le
2.0V$, which is discretized onto grids with $N_x=N_y=N_z=64$ and
$N^{\rm v}_x=N^{\rm v}_y=N^{\rm v}_z=64$ or $32$. Note that the
extension of the velocity space is larger than in the previous test,
because some portion of the matter can have large velocities during
the merging of the two spheres.

For comparison, we perform an $N$-body simulation of the same system.
The initial conditions are set up in the same manner except that each
King sphere is represented by $10^6$ particles. In this $N$-body
simulation, we adopt the Particle-Mesh (PM) method as a Poisson
solver, in which the triangular-shaped cloud (TSC) mass assignment
scheme is used in computing mass density field from the particle
distribution. The gravitational force is calculated using the 4-point
finite difference scheme.  The number of grid points to compute the
gravitational potential is set to 64 for each of $x$-, $y$- and
$z$-dimension, which gives effectively the same spatial resolution as
that of the Vlasov--Poisson solver.

\begin{figure}[htbp]
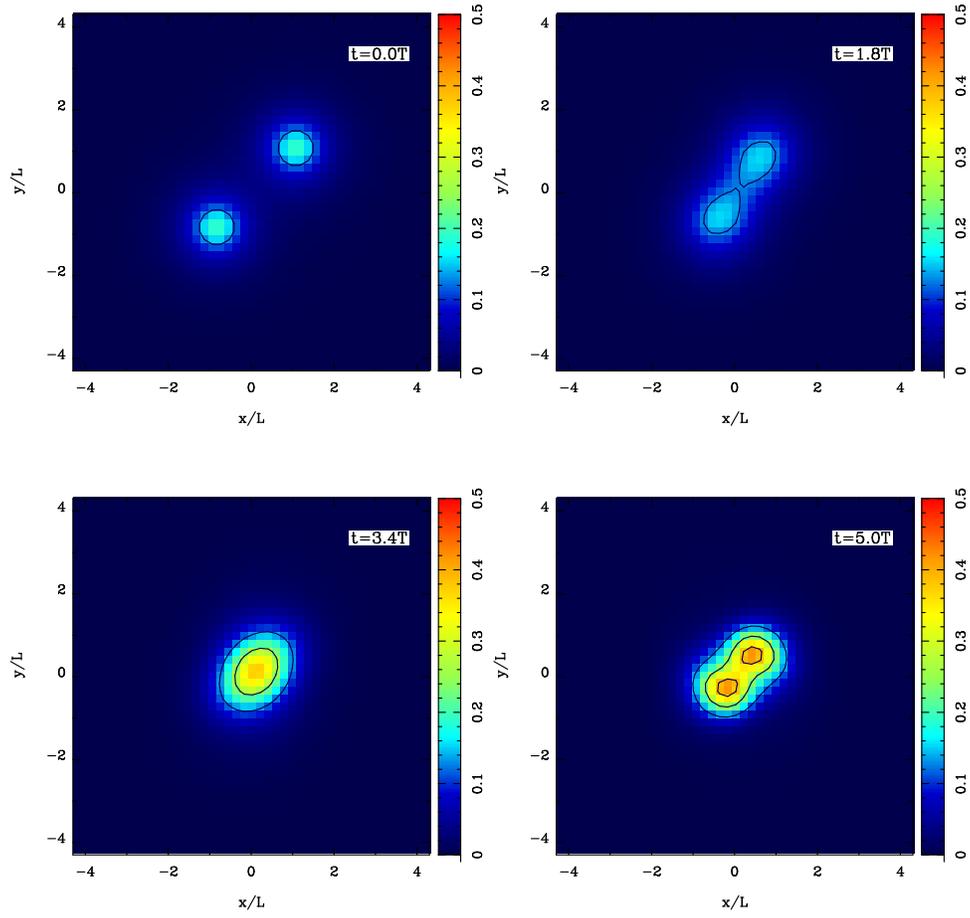

 \leavevmode
 \begin{center}
  \includegraphics[width=6.3cm]{king_merging.init.eps}
  \includegraphics[width=6.3cm]{king_merging-04.eps}
  \includegraphics[width=6.3cm]{king_merging-08.eps}
  \includegraphics[width=6.3cm]{king_merging-12.eps}

  \figcaption{Test-6: Maps of the mass density $\rho(\itbold{x})$ on
    the $z=0$ plane at $t=0.0$, $1.8T$, $3.4T$ and $5.0T$. The color
    scales indicate the mass density in units of
    $M/r_0^3$.\label{fig:king_merging_map}}
 \end{center}
\end{figure}

\begin{figure}[htbp]
 \leavevmode
 \begin{center}
  \includegraphics[width=6.3cm]{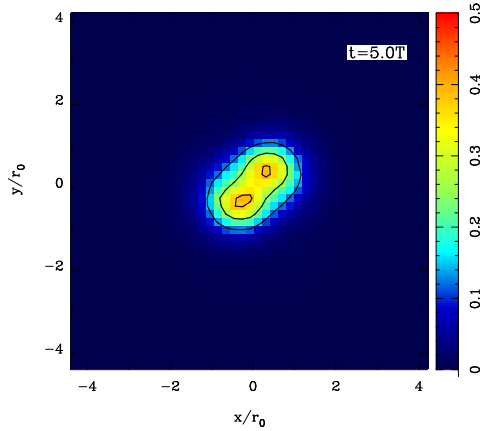}
  
  \figcaption{Test-6: A density map of the result at a $z=0$ plane from
  the $N$-body simulation at $t=5.0T$. \label{fig:king_merging_nbody_map}}
  \end{center}
\end{figure}

Figure~\ref{fig:king_merging_map} depicts the time evolution of mass
density map at $z=0$ plane in the run with $N^{\rm v}_x=N^{\rm
  v}_y=N^{\rm v}_z=64$. The cores of the two King sphere first
encounter at $t=3.4T$ and then go through each other in a
collisionless manner. The density distribution is smooth at all the
time.  We compare the mass density distribution at $t=5.0T$ between
the Vlasov--Poisson simulation and the $N$-body simulation in
Figure~\ref{fig:king_merging_map} and
Figure~\ref{fig:king_merging_nbody_map}.  Both the simulations
produce fairly consistent results, although the density distribution
in the $N$-body simulations appears slightly asymmetric between the
two spheres.

\begin{figure}[htbp]
 \leavevmode
  \begin{center}
   \includegraphics[width=13cm]{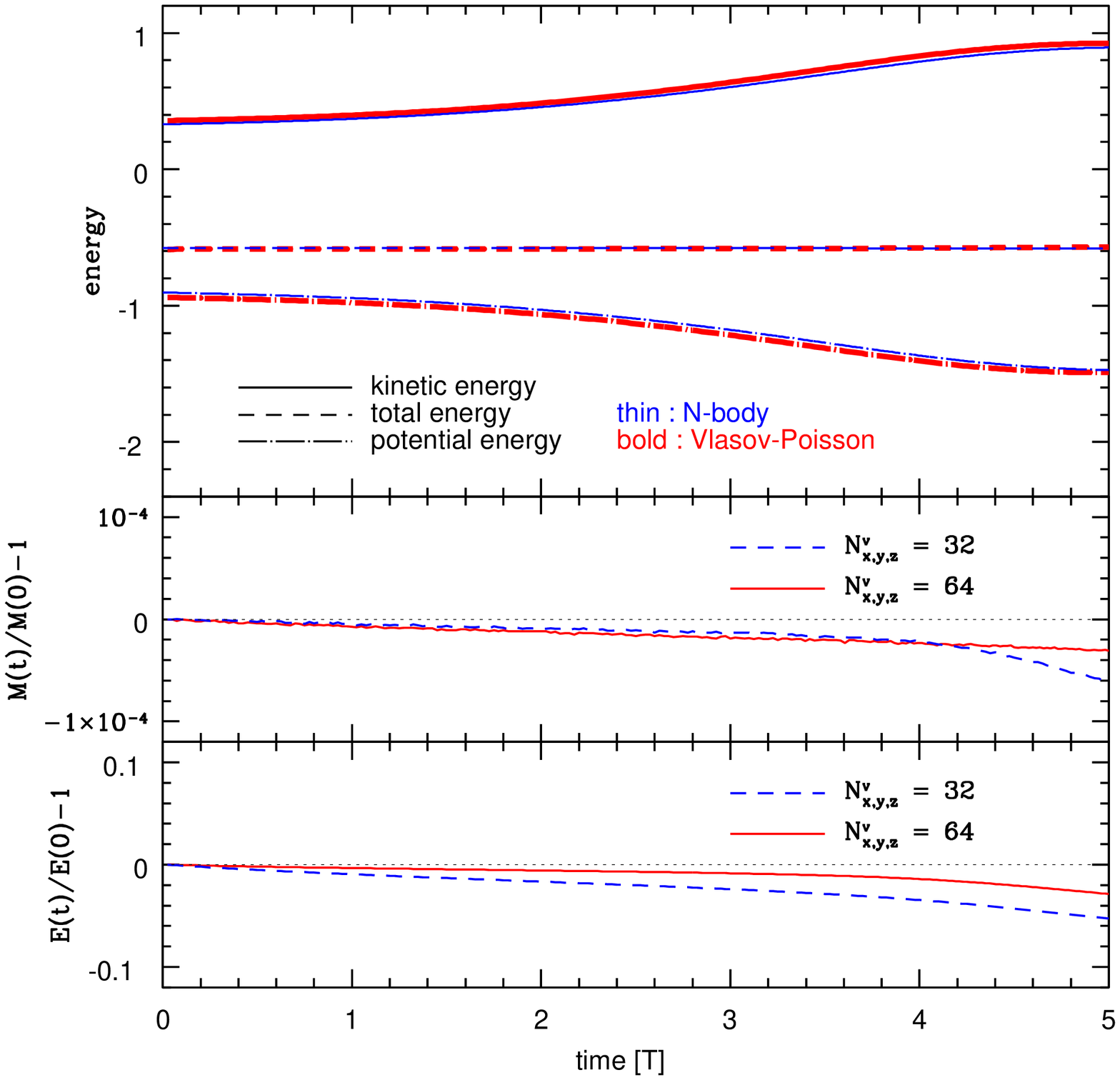} 
   \figcaption{Test 6: Time evolution of the kinetic, gravitational
   potential and total energy in the run with $N_x=N_y=N_z=N^{\rm
   v}_x=N^{\rm v}_y=N^{\rm v}_z=64$ are shown by thick lines. Those in
   the $N$-body simulation are also shown by thin lines.  The relative
   differences of the total energy and the total mass in the runs with
   $N^{\rm v}_x=N^{\rm v}_y=N^{\rm v}_z=64$ (solid lines) and $N^{\rm
   v}_x=N^{\rm v}_y=N^{\rm v}_z=32$ (dashed lines) are shown in the
   middle and bottom panels,
   respectively. \label{fig:king_merging_diag}}
  \end{center}
\end{figure}

For a more quantitative comparison between the Vlasov--Poisson and
$N$-body simulations, the time evolution of the kinetic, the
gravitational potential, and the total energies in both the
simulations are shown in the top panel of
Figure~\ref{fig:king_merging_diag}. One can see clearly the consistent
behaviors of the kinetic and gravitational potential energies in the
Vlasov--Poisson and $N$-body simulations.  The slight differences
between the two runs are primarily due to the different discretization
of the system. Relative errors of the total mass and energy
conservation in the runs with $N^{\rm v}_x=N^{\rm v}_y=N^{\rm v}_z=64$
and $N^{\rm v}_x=N^{\rm v}_y=N^{\rm v}_z=32$ are shown in
Figure~\ref{fig:king_merging_diag}.  The total mass is well conserved
with a sufficiently small relative error of $\ll 10^{-4}$ in both
resolutions in the velocity space (bottom panel).  We find a slight
decrease in the total mass at $t\gtrsim 4.5T$ in the run with the
lower velocity resolution. This is because the extent of the matter
distribution in the velocity space exceeds the predefined velocity
ranges during the merging of the cores of the two king spheres. In the
run with the higher velocity resolution, while it is also the case
that the extent of the matter distribution in the velocity space is
not fully enclosed, the deviation from the total mass conservation is
kept relatively small because the better velocity resolution enables
better reconstruction of the distribution function at the high
velocity tails.

Total energy conservation is assured better in the run with the higher
velocity resolution.  Even with the higher resolution, however, the
relative error in the total energy is $\simeq 3\%$ at $t=5T$, while it
is $\lesssim$ 1.5\% at $t<4T$. Again, this can be understood by the
fact that the extent of the matter distribution in the velocity space
is beyond the predefined velocity ranges during $4T\lesssim t \lesssim
5T$. Such velocity 'overflow' gives a stronger impact to the total
energy budget rather than the total mass conservation because matter
with a large velocity has naturally a large kinetic energy.  It would
be desirable to develop a scheme which adaptively rescales the
velocity space with proper reconstruction or re-mapping of the
velocity distribution function.

\begin{figure}[htbp]
 \leavevmode
 \begin{center}
  \includegraphics[width=8cm]{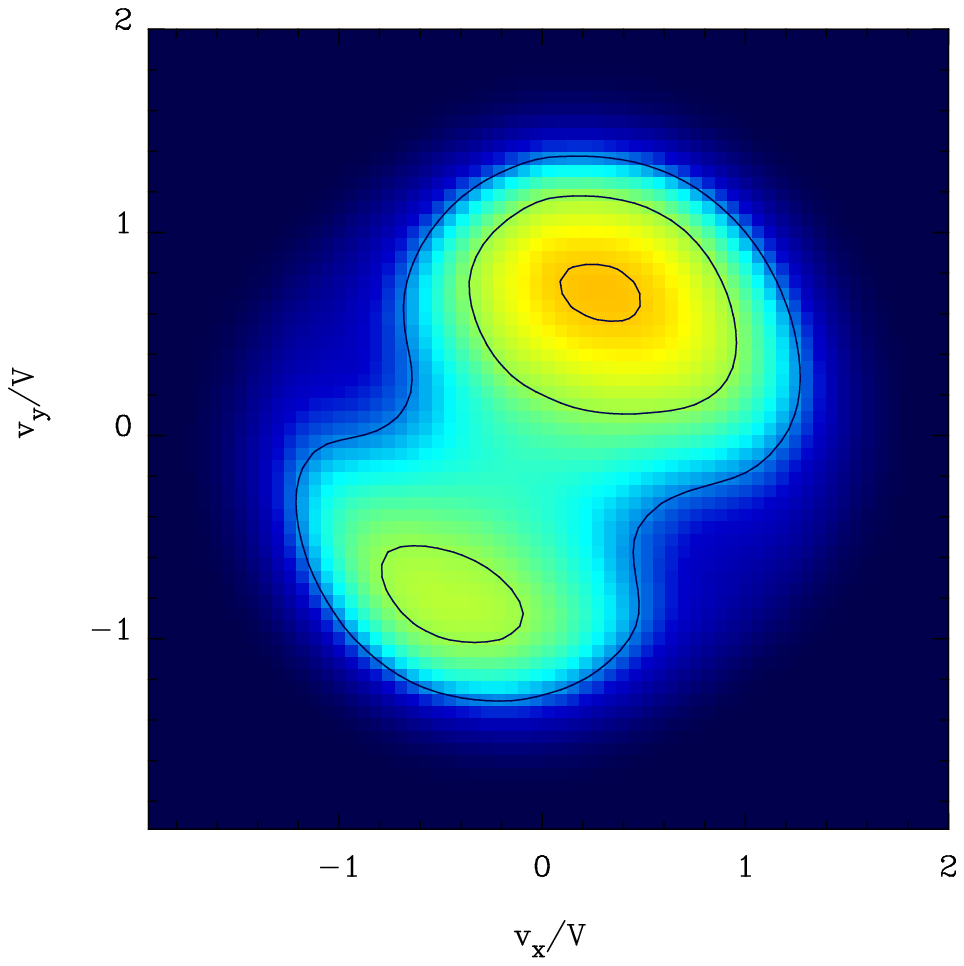}
  \includegraphics[width=8cm]{vel_space_nbody_scatter.eps} 

  \figcaption{Test 6: We compare the phase space density in the velocity
    space at a single spatial grid point near the center of the system
    at $t=4.2T$ in the Vlasov--Poisson 
    simulation (left panel) and the distribution of the particles
    in the same spatial volume in the $N$-body simulation (right panel).
    In the right panel, contours of the particle distribution 
    are also drawn
    with the same binning as the velocity grid in the Vlasov--Poisson
    simulation.
    \label{fig:vel_space_map}}
 \end{center}
\end{figure}

We have seen in Figure~\ref{fig:king_merging_map} and 
Figure~\ref{fig:king_merging_nbody_map}
that the density distributions are quite similar between the
Vlasov run and the $N$-body run.
It is interesting to compare the matter distribution in the velocity
space between the two simulations. The left
panel of Figure~\ref{fig:vel_space_map} shows the phase space density
in the velocity space at a single spatial grid point near the mass
center of the two King spheres. For this plot, 
we use the output at $t=4.2T$, when the two peaks of the phase space
density match the bulk velocities of the two King
spheres. The velocity distribution of the particles in the same
spatial volume of the $N$-body simulation are depicted in the right
panel of Figure~\ref{fig:vel_space_map}. Although there are two
broad clumps at roughly the same locations as those in the Vlasov--Poisson
run, one can clearly see severe contaminations by the shot noise. 
The velocity struture is not well sampled even in the 
$N=10^{6}$ run.

\begin{figure}[htbp]
 \begin{center}
  \includegraphics[width=8cm]{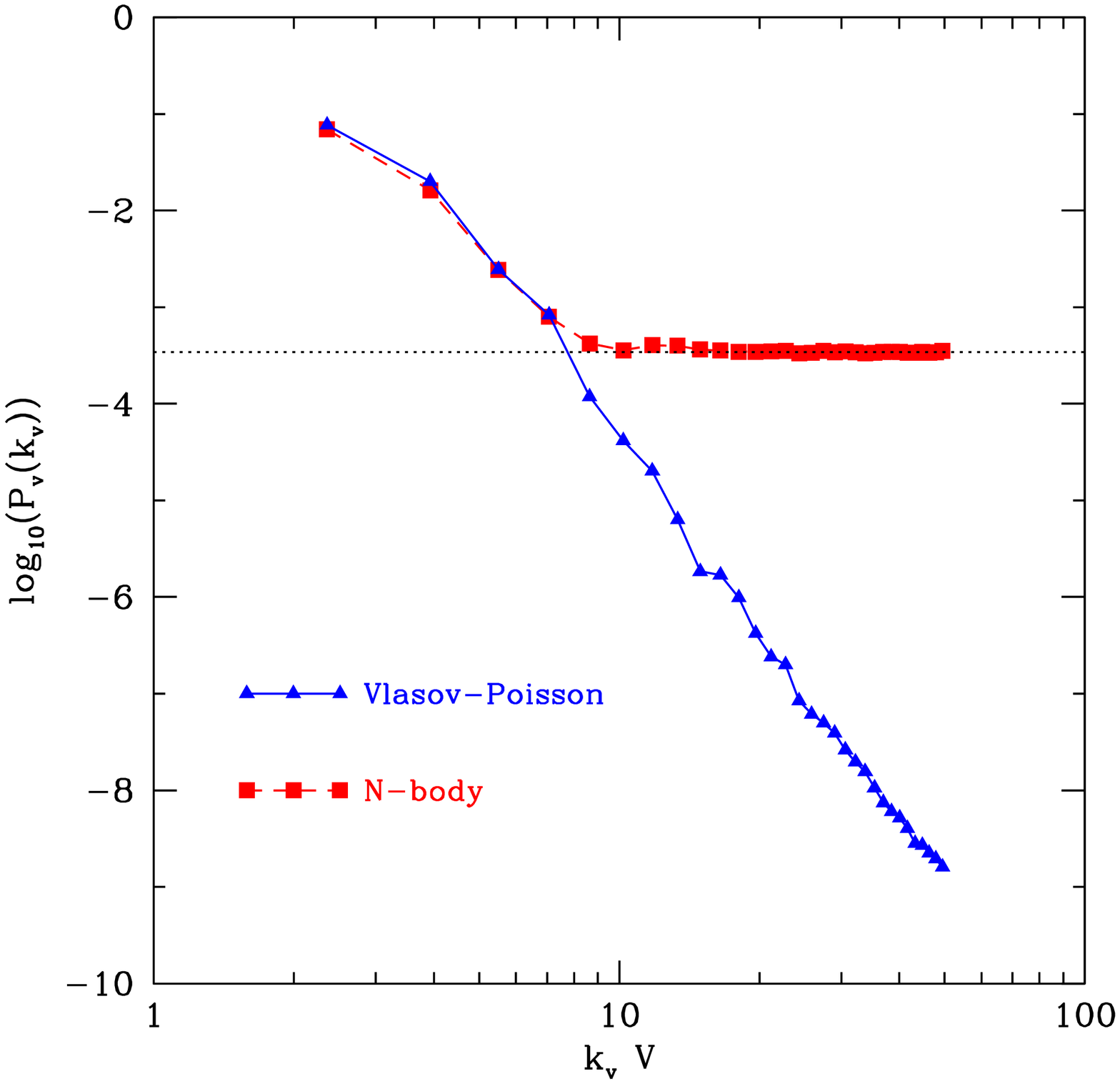} 
  
  \figcaption{Test 6: Power spectra of phase space density in the
  velocity space in the Vlasov--Poisson and $N$-body
  simulations.\label{fig:vel_power_spec}}
 \end{center}
\end{figure}

In order to quantify the shot
noise level in the velocity distribution of the $N$-body simulation,
we compute the power spectra of the velocity distribution function
\begin{equation}
  P_{\rm v} (k_{\rm v}) = \langle | \hat{F}(\itbold{k}_{\rm v}) |^2 \rangle,
\end{equation}
where $\hat{F}(\itbold{k}_{\rm v})$ is a discrete Fourier transform of
the distribution function in the velocity space and $\itbold{k}_{\rm
  v}$ is a wave number vector corresponding to a certain velocity
vector.  The velocity power spectra thus calculated at the
same spatial position as in Figure~\ref{fig:vel_space_map} are shown in
Figure~\ref{fig:vel_power_spec}. 
The two power spectra are in good agreement with each
other at large velocity scales, $k_{\rm v} V \lesssim 10$. 
At small velocity scales ($k_{\rm v}V \gtrsim 10$), however, the $N$-body
simulation exhibits a flat spectrum, showing good contrast
with the nearly power-law spectrum in the Vlasov--Poisson simulation.
The velocity power for the $N$-body run  
is significantly contaminated by the shot noise. 
On the assumption that the velocity power of the 
Vlasov--Poisson simulation is accurate to $k_{\rm v}V \sim 50$,
we argue that the same result would be obtained if we could employ
nearly a five orders-of-magnitude larger number of particles
in the $N$-body simulation. 

\section{Memory Consumption, CPU Timing and Parallelization Efficiency}

All of the simulations presented in the present paper were performed with a
large scale massively parallel supercomputer, T2K-Tsukuba system
installed at Center for Computational Sciences, University of
Tsukuba\footnote{http://www.open-supercomputer.org/}. 
Each computational node of the T2K-Tsukuba system consists of
four sockets of 2.3GHz quad-core AMD Opteron and 32GByte of DDR2 SDRAM
memory. All the nodes are connected through quad-rail of DDR
Infiniband interconnection network. 

The required memory $M$ is approximately computed as
\begin{equation}
 M = 256 \left(\frac{N^{\rm p}}{64^3}\right) \left(\frac{N^{\rm v}}{64^3}\right) {\rm GByte}, 
\end{equation}
where $N^{\rm p}=N_x N_y N_z$ and $N^{\rm v} = N^{\rm v}_x N^{\rm
v}_y N^{\rm v}_z$ are the numbers of grids in the spatial and velocity
spaces, respectively. Our Vlasov code uses single-precision floating point
numbers for storing the value of the distribution
function. Since each node of the T2K-Tsukuba system can store data up to
24 GBytes on its memory, for the runs with $N_x N_y N_z=64^3$ and
$N^{\rm v}_xN^{\rm v}_y N^{\rm v}_z=64^3$, we typically use 16--64
nodes. 

Table~\ref{tab:breakdown} shows a breakdown of the wall clock time
consumed by several parts in our code over a single timestep integration. 
All of the runs (A, B and C) are performed with 64 MPI processes and
each MPI process is also parallelized in a multi-thread manner using
the OpenMP implementation. For example, Run A adopts 16 nodes
(equivalently 256 CPU cores), and each MPI process invokes 4 threads.
Although the explored parameter space is limited, we confirm that our
code performs well on up to 1024 CPU cores (64 nodes).  Comparing Run A
and Run B, we see a good weak scaling in which the wall clock time
$T^{\rm total}$ almost precisely scales with the simulation size
($N^{\rm p} N^{\rm v}$) with the same number of nodes. On the other
hand, Run C, using four times more nodes, is only three times faster
than Run B for the same number of grids. This is because the PFC
scheme needs global maximum
values of the phase space density over the entire 1-dimensional
computational regions to ensure the maximum principle, and thus
solving the advection equations in the position space requires data
transfer among the nodes associated with the adjacent computational
regions through the inter-node network. As a result, as we see in the
difference between $T^{\rm p}$ and $T^{\rm v}$ of
table~\ref{tab:breakdown}, the advection operations in the position
space take longer than in the velocity space. The necessary data
transfer hampers the strong scaling in solving the advection equation
in the position space.

\begin{table}
 \begin{center}
  \caption{Breakdown of the wall clock time for a single time step
  \label{tab:breakdown}}
  \begin{tabular}{ccccccccc}
   \tableline\tableline
   ID & $N^{\rm p}$ & $N^{\rm v}$ & $N^{\rm node}$  & $T^{\rm
   p}$ [sec]\tablenotemark{\it a} &
   $T^{\rm v}$ [sec]\tablenotemark{\it a} & $T^{\rm grav}$
   [sec]\tablenotemark{\it b} & $T^{\rm comm}$ [sec]\tablenotemark{\it c}&
   $T^{\rm total}$ [sec]\tablenotemark{\it d}\\
   \tableline
   A & $64^3$ & $32^3$ & 16 & 9.1 & 6.4 & 1.6 & 1.95 & 72.1 \\
   B & $64^3$ & $64^3$ & 16 & 60.1 & 56.3 & 4.7 & 11.1 & 550.2 \\
   C & $64^3$ & $64^3$ & 64 & 21.2 & 14.4 & 5.1 & 10.5 & 181.4 \\
   \tableline
  \end{tabular}
  \tablenotetext{a}{Time for solving advection equations along a single
  dimension of the position and velocity spaces.}
  \tablenotetext{b}{Time for solving the Poisson equation including the
  calculation of the density field and the communication among nodes.}
  \tablenotetext{c}{Overhead for communicating the data in the adjacent
  computational subdomains.}  
  \tablenotetext{d}{Total wall clock time to advance the system by a
  single timestep.}
  \tablecomments{Since we perform three and six advection operations in
  a single timestep in the spatial and velocity grids (see
  equation~(\ref{eq:integration})), their contributions to $T^{\rm
  total}$ are $3T^{\rm p}$ and $6T^{\rm v}$, respectively.}
 \end{center}
\end{table}

\section{Summary and Discussion}

In this paper, we have developed a fully parallelized Vlasov--Poisson
solver in six-dimensional phase space for collisionless self-gravitating
systems. The Vlasov solver is based on the recently proposed positive
flux conservation scheme, whereas the Poisson solver utilizes the
conventional convolution method based on the discrete Fourier transform.
We have conducted large simulations of collisionless self-gravitating
systems on the phase space discretized onto $64^{6}$ grids.  We
have performed a suite of test calculations to examine the accuracy and
performance of our simulation code.

The results of the test suite are summarized as follows. In Test 1, we
examine the overall accuracy of the PFC scheme to solve a
1-dimensional advection equation which is adopted in all the
simulations presented in this paper. The mass and the energy
conservations are confirmed to an accuracy of $10^{-5}$ for the
one-dimensional advection problem. The initial distribution function
is well-preserved, without significant smearing due to numerical
diffusion.  In 1D and 3D tests for the time evolution of the density
perturbation through gravatational interactions (Test 2 and 4,
respectively), the growth and damping rates of the density
perturbations are consistent with the linear theory prediction at
early phases. The Galilean invariance is also explicitly shown (Test
3). In Test 5, a stable spherical solution of the Vlasov--Poisson
equations, the King sphere, is also reproduced in full six-dimensional
phase space. The results manifest that our time-integration scheme is
accurate.  Finally, our code works efficiently on massively parallel
computers. It runs well on up to 1024 CPU cores and scales well with the
problem size and with the number of processors.

We summarize the advantages of the simulations of collisionless
self-gravitating systems based on the Vlasov--Poisson equations over
the conventional $N$-body simultaions as follows.  Since the matter
distribution in the velocity space is explicitly represented in the
form of a continuum distribution function, physical processes that are
sensitive to the velocity perturbations such as Landau damping can be
treated accurately as seen in Test-2 and 4.  The collisionless feature
is assured in the Vlasov--Poisson simulations, while artificial
two-body relaxation could compromise the results of $N$-body
simulations.  The resolution in the velocity space in the
Vlasov--Poisson simulations is shown to be significantly better than
that of $N$-body simulations in which the particle distribution in the
velocity space is intrinsically rather noisy.  Currently the spatial
resolution of the Vlasov--Poisson simulations is not as impressive as
those of the state-of-the-art $N$-body simulations. However, the
performance of our grid-based Vlasov solver scales well with the
number of processors.  Thus we expect the simulation size can be
steadily increased as the available computing power increases in the
near future. We foresee direct integration of the collisionless
Boltzmann equation will be a promising method in the era of exa-flops
computing.

Further improvements of the Vlasov--Poisson solver we have developed
includes an adaptive mesh approach to improve the spatial and velocity
resolutions without significantly increasing the required amount of
the memory, and an adoption of more sophisticated schemes to solve
one-dimensional advection equations to reduce numerical errors caused
by the coarse-grained discretization of the phase space.

\section*{Acknowledgement}
We thank Kojiro Suzuki for discussions and comments.  This work is
supported in part by Grant-in-Aid for Challenging Exploratory Research
(21654026) from JSPS. NY is grateful for financial support from
Grant-in-Aid for Young Scientists (S) (20674003) and by the FIRST
program Subaru Measurements of Images and Redshifts (SuMIRe) by the
Council for Science and Technology Policy. MU is grateful to JSPS
Grant-in-Aid for Scientific Research (S) (20224002). Numerical
Simulations for this work have been carried out under the
``Interdisciplinary Computational Science Program'' in Center for
Computational Sciences, University of Tsukuba.

\end{document}